\documentstyle[preprint,psfig,tighten,eqsecnum,floats,aps]{revtex}

\def\ut#1{\rlap{\lower1ex\hbox{$\sim$}}{#1}}

\def\U{{\rm U}}
\def\SU{{\rm SU}}
\def\su{{\rm su}}
\def\s{\sigma_{a}^{AA'}}
\def\a{A_{a}^{AB}}
\def\SL{{\rm SL}}
\def\sl{{\rm sl}}

\def\D{{\cal D}}

\def\Eg{{}^\gamma\!\Sigma}

\def\i{\gamma}

\def\bA{{\bf A}}
\def\bF{{\bf F}}
\def\bE{{\bf E}}

\def\M{{\cal M}}

\def\pb#1{\rlap{\lower1.5ex\hbox{$\longleftarrow$}}{#1}}
\def\dpb#1{\rlap{\lower1.5ex\hbox{$\Longleftarrow$}}{#1}}
\def\spb#1{\rlap{\lower1.0ex\hbox{$\leftarrow$}}{#1}}
\def\sdpb#1{\rlap{\lower1.0ex\hbox{$\Leftarrow$}}{#1}}
\def\i{i}
\def\o{o}
\def\d{{\rm d}}
\def\TR{{\rm Tr}}
\def\bar{\overline}

\def\Eg{{}^\gamma\!\Sigma}

\def\Ag{{}^\gamma\!A}

\def\ba{\begin{eqnarray}}
\def\ea{\end{eqnarray}}
\def\be{\begin{equation}}
\def\ee{\end{equation}}

\def\={\mathrel{\widehat\mathalpha{=}}}
\def\puto#1{\rlap{\raise.5ex\hbox{\char'27}}{#1}}

\preprint{\vbox{\baselineskip=12pt
\rightline{NSF-ITP-9-33}}}

\begin{document}
\draft
\title{Isolated Horizons: the Classical Phase Space}
\author {Abhay\ Ashtekar${}^{1,3}$, 
Alejandro\ Corichi${}^{1,2,3}$
and Kirill\ Krasnov${}^{1,3}$
}
\address{1. Center for Gravitational Physics and Geometry \\
Department of Physics, The Pennsylvania State University \\
University Park, PA 16802, USA}

\address{2. Instituto de Ciencias Nucleares\\
Universidad Nacional Aut\'onoma de M\'exico\\
A. Postal 70-543, M\'exico D.F. 04510, M\'exico.}

\address{3. Institute for Theoretical Physics,\\ 
University of California, Santa Barbara, CA 93106, USA}
\maketitle

\begin{abstract}

A Hamiltonian framework is introduced to encompass non-rotating (but
possibly charged) black holes that are ``isolated'' near future
time-like infinity or for a finite time interval.  The underlying
space-times need not admit a stationary Killing field even in a
neighborhood of the horizon; rather, the physical assumption is that
neither matter fields nor gravitational radiation fall across the
portion of the horizon under consideration.  A precise notion of
non-rotating isolated horizons is formulated to capture these ideas.
With these boundary conditions, the gravitational action fails to be
differentiable unless a boundary term is added at the horizon.  The
required term turns out to be precisely the Chern-Simons action for
the self-dual connection.  The resulting symplectic structure also
acquires, in addition to the usual volume piece, a surface term which
is the Chern-Simons symplectic structure.  We show that these
modifications affect in subtle but important ways the standard
discussion of constraints, gauge and dynamics.  In companion papers,
this framework serves as the point of departure for quantization, a
statistical mechanical calculation of black hole entropy and a
derivation of laws of black hole mechanics, generalized to isolated
horizons.  It may also have applications in classical general
relativity, particularly in the investigation of of analytic issues
that arise in the numerical studies of black hole collisions.

\end{abstract}

\section{Introduction}
\label{sec1}

In the seventies, there was a flurry of activity in black hole physics
which brought out an unexpected interplay between general relativity,
quantum field theory and statistical mechanics \cite{1,2,3}.  That
analysis was carried out only in the semi-classical approximation,
i.e., either in the framework of Lorentzian quantum field theories in
curved space-times or by keeping just the leading order, zero-loop
terms in Euclidean quantum gravity.  Nonetheless, since it brought
together the three pillars of fundamental physics, it is widely
believed that these results capture an essential aspect of the more
fundamental description of Nature.  For over twenty years, a concrete
challenge to all candidate quantum theories of gravity has been to
derive these results from first principles, without having to invoke
semi-classical approximations.

Specifically, the early work is based on a somewhat ad-hoc mixture of
classical and semi-classical ideas ---reminiscent of the Bohr model of
the atom--- and generally ignored the quantum nature of the
gravitational field itself.  For example, statistical mechanical
parameters were associated with macroscopic black holes as follows.
The laws of black hole mechanics were first derived in the framework
of {\it classical} general relativity, without any reference to the
Planck's constant $\hbar$ \cite{2}.  It was then noted that they have a
remarkable similarity with the laws of thermodynamics if one
identifies a multiple of the surface gravity $\kappa$ of the black
hole with temperature and a corresponding multiple of the area $a_{\rm
hor}$ of its horizon with entropy.  However, simple dimensional
considerations and thought experiments showed that the multiples must
involve $\hbar$, making quantum considerations indispensable for a
fundamental understanding of the relation between black hole mechanics
and thermodynamics \cite{1}.  Subsequently, Hawking's investigation of
(test) quantum fields propagating on a black hole geometry showed that
black holes emit thermal radiation at temperature $T_{\rm rad} =
\hbar\kappa/2\pi$ \cite{3}.  It therefore seemed natural to assume
that black holes themselves are hot and their temperature $T_{\rm bh}$
is the same as $T_{\rm rad}$.  The similarity between the two sets of
laws then naturally suggested that one associate entropy $S= a_{\rm
hor}/4\hbar$ with a black hole of area $a_{\rm hor}$.  While this
procedure seems very reasonable, these considerations can not be
regarded as providing a ``fundamental derivation'' of the
thermodynamic parameters of black holes.  The challenge is to derive
these formulas from first principles, i.e., by regarding large black
holes as statistical mechanical systems in a suitable quantum gravity
framework.

Recall the situation in familiar statistical mechanical systems such
as a gas, a magnet or a black body.  To calculate their thermodynamic
parameters such as entropy, one has to first identify the elementary
building blocks that constitute the system.  For a gas, these are
molecules; for a magnet, elementary spins; for the radiation field in
a black body, photons.  What are the analogous building blocks for
black holes?  They can not be gravitons because the gravitational
fields under consideration are stationary.  Therefore, the elementary
constituents must be non-perturbative in the field theoretic sense.
Thus, to account for entropy from first principles within a candidate
quantum gravity theory, one would have to: i) isolate these
constituents; ii) show that, for large black holes, the number of
quantum states of these constituents goes as the exponential of the
area of the event horizon; iii) account for the Hawking radiation in
terms of quantum processes involving these constituents and matter
quanta; and, iv) derive the laws of black hole thermodynamics from
quantum statistical mechanics.

These are difficult tasks, particularly because the very first step
--isolating the relevant constituents-- requires new conceptual as
well as mathematical inputs.  Furthermore, in the semi-classical
theory, thermodynamic properties have been associated not only with
black holes but also with cosmological horizons.  Therefore, the
framework has to be sufficiently general to encompass these diverse
situations.  It is only recently, more than twenty years after the
initial flurry of activity, that detailed proposals have emerged.  The
more well-known of these comes from string theory \cite{4} where the
relevant elementary constituents are associated with D-branes which
lie outside the original perturbative sector of the theory.  The
purpose of this series of articles is to develop another scenario,
which emphasizes the quantum nature of geometry, using
non-perturbative techniques from the very beginning.  Here, the
elementary constituents are the quantum excitations of geometry itself
and the Hawking process now corresponds to the conversion of the
quanta of geometry to quanta of matter.  Although the two approaches
seem to be strikingly different from one another, we will see \cite{5}
that they are in certain sense complementary.
 
An outline of ideas behind our approach was given in \cite{6,7}.  In
this paper, we will develop in detail the classical theory that
underlies our analysis. The next paper \cite {5} will be devoted to
the details of quantization and to the derivation of the entropy
formula for large black holes from statistical mechanical
considerations. A preliminary account of the black hole radiance in
this approach was given in \cite{8} and work is now in progress on
completing that analysis. A derivation of the laws governing isolated
horizons --which generalize the standard zeroth and first laws of
black hole mechanics normally proved in the stationary context-- is
given in \cite{abf1,abf2}.

The primary goal of our classical framework is to overcome three
limitations that are faced by most of the existing treatments.  First,
isolated black holes are generally represented by {\it stationary}
solutions of field equations, i.e., solutions which admit a
translational Killing vector field {\it everywhere}, not just in a
small neighborhood of the black hole.  While this simple idealization
was appropriate in the early development of the subject, it does seem
overly restrictive.  Physically, it should be sufficient to impose
boundary conditions at the horizon to ensure {\it only that the black
hole itself is isolated}.  That is, it should suffice to demand only
that the intrinsic geometry of the horizon be time independent
although the geometry outside may be dynamical and admit gravitational
and other radiation.  Indeed, we adopt a similar viewpoint in ordinary
thermodynamics; in the standard description of equilibrium
configurations of systems such as a classical gas, one usually assumes
that only the system is in equilibrium and stationary, not the whole
world.  For black holes, in realistic situations, one is typically
interested in the final stages of collapse where the black hole is
formed and has ``settled down'' (Figure 1) or in situations in which
an already formed black hole is isolated for the duration of the
experiment.  In such situations, there is likely to be gravitational
radiation and non-stationary matter far away from the black hole,
whence the space-time as a whole is not expected to be stationary.
Surely, black hole thermodynamics should be applicable in such
situations.

\begin{figure}
\centerline{
\hbox{\psfig{figure=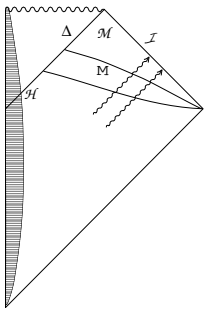,height=4cm}}}
\bigskip
\caption{A typical gravitational collapse.  The portion $\Delta$ of the
      horizon at late times is isolated.  The space-time $\mathcal{M}$
      of interest is the triangular region bounded by $\Delta$, ${\cal
      I}^+$ and a partial Cauchy slice $M$.}\label{exam}
\end{figure}

The second limitation comes from the fact that the classical framework
is generally geared to {\it event} horizons which can only be
constructed retroactively, after knowing the {\it complete} evolution
of space-time.  Consider for example, Figure 2 in which a spherical
star of mass $M$ undergoes a gravitational collapse.  The singularity
is hidden inside the null surface $\Delta_1$ at $r=2M$ which is
foliated by a family of trapped surfaces and which would be a part of
the event horizon if nothing further happens in the future.  However,
let us suppose that, after a very long time, a thin spherical shell of
mass $\delta M$ collapses.  Then, $\Delta_1$ would not be a part of
the event horizon which would actually lie slightly outside $\Delta_1$
and coincide with the surface $r= 2(M+\delta M)$ in distant future.
However, on physical grounds, it seems unreasonable to exclude
$\Delta_1$ a priori from all thermodynamical considerations.  Surely,
one should be able to establish laws of black hole mechanics not only for
the event horizon but also for $\Delta_1$.  Another example is
provided by cosmological horizons in the de Sitter space-time.  In this
space-time, there are no singularities or event horizons.  On the
other hand, semi-classical considerations enable one to assign entropy
and temperature to these horizons as well.  This suggests that the
notion of event horizons is too restrictive for thermodynamic
considerations.  We will see that this is indeed the case; as far as
equilibrium properties are concerned, the notion of event horizons can
be replaced by a more general, quasi-local notion of ``isolated
horizons'' for which the familiar laws continue to hold.  The surface
$\Delta_1$ in figure 2 as well as the cosmological horizons in
de Sitter space-times are examples of isolated horizons.

\begin{figure}
\centerline{
\hbox{\psfig{figure=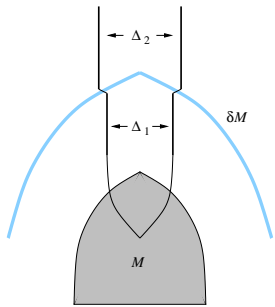,height=4cm}}}
\bigskip
\caption{A spherical star of mass $M$ undergoes collapse. Later, a
      spherical shell of mass $\delta{M}$ falls into the resulting
      black hole. While $\Delta_1$ and $\Delta_2$ are both isolated
      horizons, only $\Delta_2$ is part of the event
      horizon.}\label{shell}
\end{figure}

The third limitation is that most of the existing classical treatments
fail to provide a natural point of departure for quantization.  In a
systematic approach, one would first extract an appropriate sector of
the theory in which space-time geometries satisfy suitable conditions
at interior boundaries representing horizons, then introduce a
well-defined action principle tailored to these boundary conditions,
and finally construct the Hamiltonian framework by spelling out the
symplectic structure, constraints, gauge and dynamics. By contrast,
treatments of black hole mechanics are often based on differential
geometric identities and field equations and not concerned with issues
related to quantization.  We will see that all these steps necessary
for quantization can be carried out in the context of isolated
horizons.

At first sight, it may appear that only a small extension of the
standard framework based on stationary event horizons may be needed to
overcome these three limitations.  However, this is not the case.  For
example, if there is radiation outside the black hole, one can not
identify the ADM mass with the mass of the black hole. Hence, to
formulate the first law, a new expression of the black hole mass is
needed.  Similarly, in absence of a space-time Killing field, we need
to generalize the notion of surface gravity in a non-trivial fashion.
Indeed, even if the space-time happens to be static but only in a
neighborhood of the horizon --already a stronger condition than what
is contemplated above-- the notion of surface gravity is ambiguous
because the standard expression fails to be invariant under constant
rescalings of the Killing field. When a {\it global} Killing field
exists, the ambiguity is removed by requiring that the Killing field
be unit at {\it infinity}. Thus, contrary to what one would
intuitively expect, the standard notion of surface gravity of a
stationary black hole refers not just to the structure at the horizon
but also that at infinity.  This ``normalization problem'' in the
definition of surface gravity seems difficult especially in the
cosmological context where Cauchy surfaces are compact.  Apart from
these conceptual problems, a host of technical issues need to be
resolved because, in the Einstein-Maxwell theory, while the space of
stationary solutions with event horizons is finite dimensional, the
space of solutions admitting isolated horizons is {\it infinite}
dimensional, since these solutions can admit radiation near infinity.
As a result, the introduction of a well-defined action principle is
subtle and the Hamiltonian framework acquires certain qualitatively
new features.

The organization of this paper is as follows.  Section \ref{sec2}
recalls the formulation of general relativity in terms of
$\SL(2,C)$-spin soldering forms and self dual connections for
asymptotically flat space-times {\it without} internal boundaries.
Section \ref{sec3} specifies the boundary conditions that define for
us isolated, non-rotating horizons and discusses those consequences of
these boundary conditions that are needed in the Hamiltonian
formulation and quantization.  It turns out that the usual action
(with its boundary term at infinity) is not functionally
differentiable in presence of isolated horizons.  However, one can add
to it a Chern-Simons term at the horizon to make it differentiable.
This unexpected interplay between general relativity and Chern-Simons
theory is discussed in Section \ref{sec4}.  It also contains a
discussion of the resulting phase space, symplectic structure,
constraints and gauge on which quantization \cite{5} will be based.
For simplicity, up to this point, the entire discussion refers to
vacuum general relativity.  The modifications that are necessary for
incorporating electro-magnetic and dilatonic hair are discussed
in Section \ref{sec5}.  Section \ref{sec6} summarizes the main
results.

Some of our constructions and results are similar to those that have
appeared in the literature in different contexts. In particular, the
ideas introduced in Section \ref{sec3} are closely related to those
introduced by Hayward in an interesting series of papers \cite{9}
which also aims at providing a more physical framework for discussing
black holes. Our introduction in Section \ref{sec4} of the boundary
term in the action uses the same logic as in the work of Regge and
Teitelboim \cite{10}, Hawking and Gibbons \cite{11} and others. The
specific form of the boundary term is the same as that in the work of
Momen \cite{12}, Balachandran et al \cite{13}, and Smolin
\cite{ls}. However, the procedure used to arrive at the term and its
physical and mathematical role are quite different. These similarities
and differences are discussed at the appropriate places in the text.

\section{Preliminaries: Review of connection dynamics}
\label{sec2}

In this paper we use a formulation of general relativity \cite{16}
in which it is a dynamical theory of connections rather than metrics.
This shift of the point of view does not change the theory
classically%
\footnote{The shift, does suggest natural extensions of general
relativity to situations in which the metric may become degenerate.
However, in this paper we will work with standard general relativity,
i.e., assume that the metrics are non-degenerate.}
and is therefore not essential to the discussion of results that hold
in classical general relativity, such as the (generalized) laws of
black hole mechanics \cite{abf1,abf2}.  However, this shift makes the
kinematics of general relativity the same as that of $\SU(2)$
Yang-Mills theory, thereby suggesting new non-perturbative routes to
quantization.  The quantum theory, discussed in detail in the
accompanying paper \cite{5}, uses this route in an essential way.
Therefore, in this paper, we will discuss the boundary conditions,
action and the Hamiltonian framework using connection variables.  To
fix notation and to acquaint the reader with the basic ideas, in this
section we will recall some facts on the classical connection
dynamics. (For further details see, e.g., \cite{18}.) For
definiteness, we will tailor our main discussion to the case in which
the cosmological constant $\Lambda$ vanishes. Incorporation of a
non-zero $\Lambda$ only requires appropriate changes in the
boundary conditions and surface terms at infinity. The structure at 
isolated horizons will remain unchanged.

\subsection{Self-dual connections}
\label{sec2.1}

Fix a 4-dimensional manifold $\M$ with only one asymptotic region.
Our basic fields will consist of a pair $(\s, \a)$ of asymptotically
flat, smooth fields where $\s$ is a soldering form for (primed and
unprimed) $\SL(2,C)$ spinors (sometimes referred to as a `tetrad') and
$\a$ is an $\SL(2,C)$ connection on unprimed spinors%
\footnote{Here, the term ``spinors'' is used in an abstract sense 
since we do not have a fixed metric on $\M$. Thus, our spinor fields 
$\alpha^{A}$ and $\beta^{A'}$ are just cross-sections of 2-dimensional 
complex vector bundles equipped with 2-forms $\epsilon_{AB}$ and 
$\epsilon_{A'B'}$. Spinor indices are raised and lowered using these two 
forms, e.g., $\alpha_{A} = \alpha^{B} \epsilon_{BA}$. For details, 
see, e.g., \cite{18,pr} and appendix A.}
\cite{18}.  Each pair $(\s,\a)$ represents a possible history. The
action is given by \cite{17}
\begin{eqnarray}
\label{action}
S'_{\rm Grav} [\sigma,A] &=& -{i\over 8\pi G}\,\,\left[\int_{\cal M}\,
{\rm Tr}\left(\Sigma\wedge F\right) - \int_{\cal T} {\rm Tr}
\left(\Sigma\wedge A \right) \right] \\ 
&\equiv& {i\over 8\pi G}\, {1\over4}\, \left[
\int_{\cal M} \d^4x \,\,\Sigma_{ab}^{AB}\,F_{cd AB}\,
\eta^{abcd} - 2\int_{\cal \tau} \d^3x\,\, \Sigma_{ab}^{AB}\, A_{cAB}\,
\eta^{abc}\right]
\nonumber
\end{eqnarray}
Here ${\cal T}$ is the time-like cylinder at infinity, the 2-forms
$\Sigma$ are given by $\Sigma^{AB} = \sigma^{AA'}\wedge
\sigma_{A'}{}^{B}$,\\ $F$ is the curvature of the connection $A$, i.e.
$F = \d A + A\wedge A$,  and $\eta$ is the (metric-independent)
Levi-Civita density. If we define a metric $g$ of signature $-+++$ via
$g_{ab} = \sigma_{a}^{AA'} \sigma_{b}^{BB'}
\epsilon_{AB}\epsilon_{A'B'}$, then the 2-forms $\Sigma$ are
self-dual (see \cite{18}, Appendix A.):
\begin{equation}
{}^\star\Sigma_{cd}{}^{AB} :=
{1\over 2} \varepsilon^{ab}{}_{cd}\Sigma_{ab}{}^{AB}=i\,
\Sigma_{cd}{}^{AB},
\label{self-dual}
\end{equation}
where $\varepsilon$ is the natural volume 4-form defined by the
metric $g$.

The equations of motion follow from variation of the action
(\ref{action}).  Varying (\ref{action}) with respect to
$A$ one gets
\be \label{eom1}
\D \wedge\Sigma =0. 
\ee
This equation implies that the connection $D$ defined by $A$ coincides
with the restriction to unprimed spinors of the torsion-free
connection $\nabla$ defined by the soldering form $\sigma$ via
$\nabla_a \sigma_b^{BB'}= 0$. The connection $\nabla$ acts on tensor
fields as well as primed and unprimed spinor fields. Thus, $A$ is now
the self-dual part of the spin-connection compatible with $\sigma$
\cite{pr}. Hence, there is a relation between the curvature $F$ and
the Riemann curvature of the metric $g$ determined by $\sigma$: $F$ is
the self-dual part of the Riemann tensor:
\begin{equation} \label{1}
F_{ab}^{AB} =  - {1\over 4}\, \Sigma_{cd}{}^{AB}\, R_{ab}{}^{cd}\, .
\end{equation}
(see, e.g., \cite{18}, p. 292).  Using this expression of the
2-form $F$ in (\ref{action}) one finds that the volume term in the
action reduces precisely to the Einstein-Hilbert term:
\begin{equation}\label{EHaction}
{1\over 16\pi G}\int_{\cal M} \d^4x \sqrt{-g} R\, .
\end{equation}
The numerical coefficients in (\ref{action}) were chosen to ensure this
precise reduction.

Varying the action (\ref{action}) with respect to $\sigma$ and taking
into account the fact that $A$ is compatible with $\sigma$, we obtain
a second equation of motion
\be \label{eom2}
G_{ab}= 0
\ee
where $G$ is the Einstein tensor of $g$. Thus, the equations of motion
that follow from the action (\ref{action}) are the same as those that
follow from the usual Einstein-Hilbert action; the two theories are
classically equivalent. 

We are now ready to perform a Legendre transform to pass to the
Hamiltonian description. For this, we now assume that the space-time
manifold $\M$ is topologically $M\times R$ for some 3-manifold $M$,
with no internal boundaries and a single asymptotic region. The basic
phase-space variables turn out to be simply the pull-backs to $M$ of
the connection $A$ and the 2-forms $\Sigma$. To avoid proliferation
of symbols, we will use the same notation for the four-dimensional
fields and their pull-backs to $M$; the context will make it clear
which of the two sets we are referring to.

The geometrical meaning of these phase-space variables is as
follows. Recall first that the four-dimensional $\SL(2,C)$-soldering
form on $\M$ induces on $M$ a 3-dimensional $\SU(2)$ soldering form
$\sigma^{a}_{AB}$ for $\SU(2)$ ``space-spinors'' on $M$ via
$$\sigma^{a}_{AB} = i\sqrt{2} q^{a}_{m} \sigma^{m}_{AA'}
\tau^{A'}{}_{B},$$ 
where $q^{a}_{m}$ is the projection operator of $M$
and $\tau^{AA'}= \tau^{a}\sigma_{a}^{AA'}$ is the spinorial
representation of the future directed, unit normal $\tau^{a}$ to
$M$. The intrinsic metric $q_{ab}$ on $M$ can be expressed as
$q_{ab} = \sigma_a^{AB} \sigma_{b\, AB}$. The 2-forms $\Sigma$, 
pulled-back to $M$, are closely related to the dual of
these $\SU(2)$ soldering forms. More precisely, 
$$ {1\over 2\sqrt{2}}\,\, \eta^{abc}\, \Sigma_{bc\, AB} =:  
\tilde\sigma_a^{AB} \equiv \sqrt{q}\, \sigma_a^{AB}$$
where $q$ is the determinant of the 3-metric $q_{ab}$ on $M$.  To see
the geometric meaning of $A$, recall first that the $\SU(2)$ soldering
form $\sigma$ determines a unique torsion-free derivative operator on
(tensor and) spinor fields on $M$.  Denote the corresponding
spin-connection by $\Gamma_{a}^{AB}$. Then, assuming that the
compatibility condition (\ref{eom1}) holds, difference between the
(pulled-back) connection $A$ and $\Gamma$ is given by the extrinsic
curvature $K_{ab}$ of $M$: 
$$A_{a}^{AB} = \Gamma_{a}^{AB} - {i\over\sqrt{2}} K_{a}^{AB}\, ,$$
where $K_{a}^{AB} = K_{ab} \sigma^{b\, AB}$. (The awkward factors of
$\sqrt{2}$ here and related formulas in Sections \ref{sec2.2} and
\ref{sec4.3} disappear if one works in the adjoint rather than the
fundamental representation of $\SU(2)$. See \cite{18}, chapter 5.)
Thus, while $\Gamma$ depends only on the spatial derivatives of the
$\SU(2)$ soldering form $\sigma$, $A$ depends on both spatial and
temporal derivatives.

The phase space consists of pairs $(A_{a}^{AB}, \Sigma_{ab}^{AB})$ of
smooth fields on the 3-manifold $M$ subject to asymptotic conditions
which are induced by the asymptotic behavior of fields on $\M$. These
require that the pair of fields be asymptotically flat on $M$ in the
following sense. To begin with, let us fix an $\SU(2)$ soldering form
$\puto{\sigma}$ on $M$ such that the 3-metric $\puto{q}_{ab} =
\puto{\sigma}_{a}^{AB} \puto{\sigma}_{bAB}$ it determines is flat
outside of a compact set. Then the dynamical fields $(\Sigma, A)$ are
required to satisfy:
\begin{eqnarray} \label{bc1}
\Sigma_{ab}= \left( 1 + {M(\theta, \phi)\over r} \right)
\puto{\Sigma}_{ab} + O\left({1\over r^2}\right) \nonumber \\
\TR(\puto{\sigma}^a A_a) = O\left({1\over r^3}\right), \qquad
A_a + {1\over 3} \TR(\puto{\sigma}^m A_m) \puto{\sigma}_a =
O\left({1\over r^2}\right),
\end{eqnarray}
where $r$ is the radial coordinate defined by the flat metric
 $\puto{q}_{ab}$.

Using the ``covariant phase space formalism'', one can use a standard
procedure to obtain a symplectic one-form from the action and take 
its curl to arrive at the symplectic structure on the phase space. The 
result is:
\begin{equation}\label{sym-compl}
\Omega|_{(A,\Sigma)} (\delta_1, \delta_2) =
-{i\over 8\pi G} \int_M {\rm Tr}
\left [ \delta_1 A\wedge\delta_2 \Sigma - \delta_2 A\wedge\delta_1\Sigma 
\right ],
\end{equation}
where $\delta \equiv (\delta A, \delta \Sigma)$ denotes a tangent
vector to the phase space at the point $(A, \Sigma)$. Note that,
although the action has a boundary term at infinity, the symplectic
structure does not.

This completes the specification of the phase space. On this space, 
the 3+1 form of Einstein's equations is especially simple; all 
equations are low order polynomials in the basic phase space
variables. Laws governing isolated horizons in classical general 
relativity are discussed within this framework in \cite{abf2}.

\subsection{Real connections}
\label{sec2.2} 

For passage to quantum theory, however, this framework is not as
suitable. To see this, note first that since we wish to borrow
techniques from Yang-Mills theories, it is natural to use connections
$A$ as the configuration variables and let the quantum states
be represented by suitable functions of (possibly generalized)
connections.  Then, the development of quantum theory would
require tools from functional analysis on the space of
connections. Moreover, to maintain diffeomorphism covariance, this
analysis should be carried out without recourse to a background
structure such as a metric. However, as noted above, the connections
$A$ are complex (they take values in the Lie algebra $\sl(2,C)$ rather
than $\su(2)$) and, as matters stand, the required functional analysis
has been developed fully only on the space of real connections
\cite{ai,al,b,mm,21,20}; work is still in progress to extend the 
framework to encompass complex connections. Therefore, at this stage,
the quantization strategy that has been most successful has been to
perform a canonical transformation to manifestly real variables. Since
the primary goal of this paper is to provide a Hamiltonian description
which can serve as a platform for quantization in \cite{5}, we will
now discuss these real variables.

The expression $A = \Gamma -(i/\sqrt{2})K$  of the $\SL(2,C)$ connection 
in terms of real fields $\Gamma$ and $K$ suggests an appropriate
strategy \cite{15}. For each non-zero real number $\gamma$, let
us set
\begin{eqnarray}
\Ag_a{}^{AB} := \Gamma_a{}^{AB} - {\gamma\over \sqrt{2}} K_a{}^{AB} 
\nonumber \\
\Eg_{ab}{}^{AB} := {1\over {\gamma}} \, \Sigma_{ab}{}^{AB}.
\label{realvar} 
\end{eqnarray}
It is not hard to check that variables $\Ag,\Eg$ are canonically
conjugate in the sense that the symplectic structure is given by:
\begin{equation} \label{sym2}
\Omega|_{(\Ag,\Eg)} \left(\delta_1, \delta_2\right)=
{1\over 8\pi G} \int_M {\rm Tr}
\left [ \delta_1\,\Ag\wedge\delta_2\,\Eg - 
\delta_2\,\Ag\wedge\delta_1\,\Eg \right ].
\end{equation}
where $\delta_1 \equiv (\delta_1\,\Ag,\delta_1\,\Eg)$ and 
$\delta_2 \equiv (\delta_2\,\Ag,\delta_2\,\Eg)$ are arbitrary tangent
vectors to the phase space at the point $(\Ag,\Eg)$. 

Thus, in the final picture, the real phase space is coordinatized by
manifestly real fields $(\Ag, \Eg)$ which are smooth and are subject
to the asymptotic conditions:
\begin{eqnarray} \label{bc2}
\Eg_{ab} = \frac{1}{\gamma} \left( 1 + {M(\theta, \phi)\over r} \right)
\puto{\Sigma}_{ab} + O\left({1\over r^2}\right) \nonumber \\
\TR(\puto{\sigma}^a A_a) = O\left({1\over r^3}\right), \qquad
A_a + {1\over 3} \TR(\puto{\sigma}^m A_m) \puto{\sigma}_a =
O\left({1\over r^2}\right),
\end{eqnarray}
where, as before, $r$ is the radial coordinate defined by the flat
metric $\puto{q}_{ab}$. The symplectic structure is given by
(\ref{sym2}).  Irrespective of the values of the real parameter
$\gamma$, all these Hamiltonian theories are classically
equivalent. They serve as the starting point for
non-perturbative canonical quantization. However, it turns out that
the corresponding quantum theories are unitarily {\it inequivalent}
\cite{14}. Thus, there is a quantization ambiguity and a one parameter
family of inequivalent quantum theories%
\footnote{If one succeeds in developing the  necessary steps in the 
functional calculus to carry out quantization directly using self-dual
connections, this ambiguity will presumably arise from the presence of
inequivalent measures in the construction of the quantum Hilbert 
space.},
parameterized by $\gamma$, which is referred to as the {\it Immirzi
parameter.} This ambiguity is similar to the $\theta$-ambiguity in QCD
where, again, the classical theories are equivalent for all values of
$\theta$ while the quantum theories are not \cite{JRO}.  The
inequivalence will play an important role in the next paper \cite{5}.

{\it Remark:} The original description in terms of pairs $(A, \Sigma)$
is ``hybrid'' in the sense that $\Sigma$ are ${\su}(2)$-valued 2-forms
while $A$ are $\sl(2,C)$ valued one-forms. The phase space is,
however, real. This is analogous to using $(z=q-ip, q)$ as
``coordinates'' on the real phase space of a simple harmonic
oscillator. Consequently, there are ``reality conditions'' that have
to be taken into account to eliminate the over-completeness of the
basic variables. These are trivialized in the manifestly real
description in terms of $(\Ag, \Eg)$ in the sense that now all
variables are real. However, while $A$ has a natural geometrical
interpretation --it is the pull-back to $M$ of the self-dual part of
the connection compatible with the four-dimensional $\SL(2,C)$
soldering form-- the real connections $\Ag$ do not. Indeed, their
meaning in the four dimensional, space-time setting is quite unclear.
Furthermore, the form of the Hamiltonian constraint in terms of $(\Ag,
\Eg)$ is complicated which made real variables undesirable in the
early literature. However, thanks to the more recent work of Thiemann
\cite{tt}, now this technical complication does not appear to be a
major obstruction.

In the next two sections, we will return to the original, self-dual
$\SL(2,C)$ connections $A$ and $\SL(2,C)$ soldering form $\sigma$
discussed in section \ref{sec2.1} and extend the framework outlined in
the beginning of this section to the context where there is an
internal boundary representing an isolated horizon. After casting this
extended framework in a Hamiltonian form, in section \ref{sec4.3} we
will again carry out the canonical transformation and pass to
manifestly real variables.

\section{Boundary conditions for isolated horizons}
\label{sec3}

As explained in the introduction, in this series of papers, we wish to
consider isolated horizons rather than stationary ones.  Space-times
of interest will now have an internal boundary, topologically
$S^{2}\times R$ and, as before, one asymptotic region. The internal
boundary will represent an {\it isolated, non-rotating horizon}.  (The
restriction on rotation is only for technical simplicity and we hope
to treat rotating horizons in subsequent papers.)  A typical example
is shown in Figure 1 which depicts a stellar gravitational
collapse. The space-time of interest is the wedge shaped region,
bounded by the future piece $\Delta$ of the horizon, future null
infinity ${\cal I}^{+}$ and a partial Cauchy surface extending from
the past boundary of $\Delta$ to spatial infinity $i^{o}$. In a
realistic collapse one expects emission of gravitational waves to
infinity, whence the underlying space-time can not be assumed to be
stationary.  There would be some back-scattering initially and a part
of the emitted radiation will fall in to the black hole. But one
expects, e.g., from numerical simulations, that the horizon will
``settle down'' rather quickly. In the asymptotic region near $i^{+}$,
we can assume that the part $\Delta$ of the horizon is non-dynamical
and isolated to a very good approximation; here, the area of the
horizon will be constant.

We now wish to impose on the internal boundary $\Delta$ precise
conditions which will capture these intuitive ideas. While they will
in particular incorporate isolated {\it event} horizons, as noted in
the Introduction, the conditions are quasi-local and therefore also
allow more general possibilities. All results obtained in this series
of papers ---the presence of the Chern-Simons boundary term in the
action, the Hamiltonian formulation, the derivation of generalized
laws of black hole mechanics and the calculation of entropy--- will
hold in this more general context. This strongly suggests that it is
the notion of {\it isolated horizons}, rather than event horizons of
stationary black holes that is directly relevant to the interplay
between general relativity, quantum field theory and statistical
mechanics, discovered in the seventies. For example, although there is
no black hole in the de Sitter space-time, the cosmological horizons
it admits are isolated horizons in our sense and our framework
\cite{5,abf1,abf2} leads to the Hawking temperature and entropy
normally associated with these horizons \cite{28}.

\subsection{Definition}
\label{sec3.1}

We are now ready to give the general Definition
and discuss the physical meaning and mathematical consequences of the
conditions it contains. Although the primary applications of the
framework will be to general relativity (possibly with a cosmological
constant) coupled to Maxwell and dilatonic fields, in this and the
next section we will allow general matter, subject only to the
conditions stated explicitly in the Definition.

\bigskip
{\it Definition:} The internal boundary $\Delta$ of a history
$(\M, \sigma_{a}^{AA'}, A_{a}^{AB})$ will be said to represent {\it a 
non-rotating isolated horizon} provided the following conditions hold%
\footnote{Throughout this paper, the symbol
{${\mathrel{\widehat\mathalpha{=}}}$} will stand for ``equal at points
of $\Delta$ to'', a single under-arrow will denote pull-back to
$\Delta$ and, a double under-arrow, pull-back to the preferred
2-sphere cross-sections S of $\Delta$.}:

\begin{itemize} 
\item{(i)} {\it Manifold conditions:} $\Delta$ is topologically
$S^{2}\times R$, foliated by a (preferred) family of 2-spheres S and
equipped with a direction field $l^{a}$ which is transversal to the
foliation. We will introduce a coordinate $v$ on $\Delta$ such that
$n_a :\= - D_av$ is normal to the preferred foliation.
\item{(ii)} {\it Conditions on the metric $g$ determined by $\sigma$:}
The surface $\Delta$ is null with $l^{a}$ as its null normal.
\item{(iii)} {\it Dynamical conditions:} All field equations 
hold at $\Delta$.
\item{(iv)} {\it Main conditions:} For any choice of the coordinate
$v$ on $\Delta$, let $l^a$ be so normalized that $l^an_a \= -1$. Then,
if $\o^{A}$ and $\i^{A}$ is a spinor-dyad on $\Delta$, satisfying
$\i^{A}\o_{A}\= 1$, such that $l^{a}= i\sigma^{a}_{AA'}\o^{A}
\bar{\o}^{A'}$ and $n^{a} \= i\sigma^{a}_{AA'}\i^{A}\bar{\i}^{A'}$,
then the following conditions should hold:\\ 
(iv.a) \quad $\o^{B} \spb{\D_{a}}\, \o_{B} \= 0$; and \\ 
(iv.b) \quad $\i^{B} {\spb{\D_{a}}}\, \i_{B} \= 
if(v) \bar{\pb{{m}_{a}}}$,\\ 
where $\D$ is the derivative operator defined by $A$, $f$ is a
positive function on $\Delta$ (which is constant on each 2-sphere S in
the foliation), and $\bar{m}_{a} :\= - \sigma_{a}^{AA'}\i_{A}
\bar{\o}_{A'}$ is a complex vector field tangential to the preferred
family of 2-spheres.
\item{(v)} {\it Conditions on matter:} On $\Delta$ the stress-energy
tensor of matter satisfies the following requirements:\\
(v.a) Energy condition: $T_{ab}l^b$ is a causal vector;\\
(v.b) The quantity $T_{ab}l^a n^b$ is constant on each 2-sphere S of the
preferred foliation.

\end{itemize}

Note that these conditions are imposed {\it only} at $\Delta$ and,
furthermore, the main condition involves only those geometrical fields
which are defined intrinsically on $\Delta$.  Let us first discuss the
geometrical and physical meaning of these conditions to see why they
capture the intuitive notions discussed above. The first three
conditions are rather weak and are satisfied on a variety of null
surfaces in, e.g., the Schwarzschild space-time (and indeed in any
null cone in Minkowski space). In essence it is the fourth condition
that pins down the surface as an isolated horizon.

The first condition is primarily topological. The second condition
simply asks that $\Delta$ be a null surface with $l^a$ as its null
normal. The third is a dynamical condition, completely analogous to
the one normally imposed at null infinity.  The last condition
restricts matter fields that may be present on the horizon. Condition
(v.a) is mild; it follows from the (much more restrictive) dominant
energy condition. It is satisfied by matter fields normally used in
classical general relativity, and in particular by the Maxwell and
dilatonic fields considered here. Condition (v.b) is a stronger
restriction which is used in our framework to ensure that the black
hole is non-rotating. Its specific form does seem somewhat mysterious
from a physical viewpoint.  However, we will see in section \ref{sec5}
that, in the Einstein-Maxwell-dilatonic system, this condition is
well-motivated from detailed considerations of the matter sector. In
the general context considered here, it serves to pinpoint in a
concise fashion the conditions that matter fields need to satisfy to
render the {\it gravitational part} of the action differentiable.

Let us now turn to the key conditions (iv). These conditions restrict
the pull-back to $\Delta$ of the connection $\D$ defined by $A$, or,
equivalently, of the connection $\nabla$ compatible with $\sigma$
since the dynamical condition (iii) implies that, on $\Delta$, the
action of these two operators agree on unprimed spinors. The
pull-backs to $\Delta$ are essential because the spinor fields $\i^A$
and $\o^A$ are defined only on $\Delta$. However, a subtlety arises
because there is a rescaling freedom in $\i^A$ and $\o^A$. To see
this, let us change our labeling of the preferred foliation via
$v\mapsto \tilde{v} = F(v)$ (with $F' >0$). Then, $n$ and $l$ get
rescaled and therefore also the spin-dyad: $\i^A \mapsto \tilde{\i}^A
= G^{-1} \i^A$ and $\o^A\mapsto \tilde{\o}^A= G\o^A$ where $|G|^2 =
F'(V)$. It is easy to verify that conditions (iv) continue to be
satisfied in the tilde frame with $f(v) \mapsto \tilde{f}(v) = F'(v)
f(v)$. Thus, conditions (iv) are unambiguous: If they are satisfied
for a dyad $(\i^{A}, \o^{A})$ then they are satisfied by all dyads
obtained from it by permissible rescalings.

The content of these conditions is as follows.  (iv.a) is equivalent
to asking that the null vector field $l^{a}$ be geodesic, twist-free,
expansion-free and shear-free. The first two of these properties
follow already from (ii). Furthermore, (v.a) and the Raychaudhuri
equation imply that if the mild energy condition $T_{ab}l^al^b \ge 0$
is satisfied and $l$ is expansion-free, it is also shear-free. Thus,
physically, the only new restriction imposed by (iv.a) is that $l$ be
expansion-free.  It is equivalent to the condition that the area
2-form of the 2-sphere cross-sections of $\Delta$ be constant in time
which in turn captures the idea that the horizon is
isolated. Condition (iv.b) is equivalent to asking that the vector
field $n^{a}$ is shear and twist-free, its expansion is spherically
symmetric, given by $-2f(v)$ and its Newman Penrose coefficient $\pi:=
- l^a\overline{m}^b\nabla_a n_b$ vanish on $\Delta$. These properties
imply that the isolated horizon is non-rotating. Since, furthermore,
$f$ is required to be positive, we are asking that the expansion of
the congruence $n^{a}$ be {\it negative}. This captures the idea that
we are interested in future horizons rather than past, i.e., black
holes rather than white holes. Finally, note that rather than fixing a
preferred foliation in the beginning, we could have required only that
a foliation satisfying our conditions exists.  The requirement (iv.b)
implies that the foliation is unique. Furthermore, it has a natural
geometrical meaning: since the expansion of $n$ is constant in this
foliation, it is the analog for null surfaces of the constant mean
curvature slicing often used to foliate space-times.

Let us summarize. Non-rotating, isolated horizons $\Delta$ are null
surfaces, foliated by a family of marginally trapped 2-spheres with
the property that the expansion of the inward pointing null normal
$n^a$ to the foliation is constant on each leaf and negative. The
presence of trapped surfaces motivates the label `horizon' while the
fact that they are {\it marginally} trapped ---i.e., that the
expansion of $l^a$ vanishes--- accounts for the adjective `isolated'.
The condition that the expansion of $n^{a}$ is negative says that
$\Delta$ is a future horizon rather than past and the additional
restrictions on the derivative of $n^a$ imply that $\Delta$ is
non-rotating. Boundary conditions refer only to the behavior of
fields at $\Delta$ and the general spirit is very similar to the
way one formulates boundary conditions at null infinity.

{\it Remarks:} 

a) All the boundary conditions are satisfied by static black holes in
the Einstein-Maxwell-dilaton theory possibly with a cosmological
constant. To incorporate rotating black holes, one would only have to
weaken conditions (iv.b) and (v.b); the rest of the framework will
remain unchanged. (Recent results of Lewandowski \cite{jl} show
that we can continue to require that the expansion of $n^a$ be
spherically symmetric in the rotating case.  However $n^a$ would now
have shear whence, there would be an additional term proportional to
$m_a$ on the right side of (iv.b).) Similarly, one may be primarily
interested in solutions to Einstein's equations with matter without
regard to whether the theory admits a well-defined action principle or
a Hamiltonian formulation. Then, one may in particular want to allow
matter rings and cages around the horizon. With such sources, the
horizons can be distorted even in static situations. To incorporate
such black holes, again, one would only have to weaken conditions
(iv.b) and (v.b).

b) Note however that, in the non-static context, there may well exist
physically interesting distorted black holes satisfying our
conditions. Indeed, one can solve for all our conditions and show that
the resulting 4-metrics need not be static or spherically symmetric on
$\Delta$ \cite{jl}. (We will see explicitly in Sections \ref{sec3.2}
and \ref{sec5.1} that the Weyl curvature and the Maxwell field need
not be spherically symmetric near $\Delta$.) Since the boundary
conditions allow such histories and since we are primarily interested
in histories ---or, in the Hamiltonian formulation, the full phase
space--- rather than classical solutions in this paper and its sequel
\cite{5}, we chose the adjective `non-rotating' rather than
`spherical' while referring to these isolated horizons.

c) In the choice of boundary conditions, we have tried to strike the
usual balance: On the one hand the conditions are strong enough to
enable one to prove interesting results (e.g., a well-defined action
principle, a Hamiltonian framework, and a generalization of black hole
mechanics) and, on the other hand, they are weak enough to allow a
large class of examples. As we already remarked, the standard
non-rotating black holes in the Einstein-Maxwell-dilatonic systems
satisfy these conditions. More importantly, starting with the standard
static black holes and using known existence theorems one can specify
procedures to construct new solutions to field equations which admit
isolated horizons as well as radiation at null infinity \cite{abf2}.
These examples already show that, while the standard static solutions
have only a finite parameter freedom, the space of solutions admitting
isolated horizons is {\it infinite} dimensional.  Thus, in the
Hamiltonian picture, even the reduced phase-space is infinite
dimensional; the conditions indeed admit a very large class of
examples.

\subsection{Symmetries and Gauge on $\Delta$}
\label{sec3.3}

In the bulk, the symmetry group is the group of automorphisms of the
$\SL(2,C)$ spin-bundle, i.e., the semi-direct product of local
$\SL(2,C)$ transformations on spinor fields with the diffeomorphism
group of $\M$. The boundary conditions impose restrictions on
dynamical fields and hence also on the permissible behavior of these
transformations on boundaries. The restrictions at infinity are
well-known: all transformations are required to preserve asymptotic
flatness. Usually, these boundary conditions involve fixing (a
trivialization of the spin-bundle and) a flat $\SL(2,C)$ soldering
form at infinity, imposing conditions on the fall-off of $\sigma$, $A$
and matter fields and a requirement that the magnetic part of the Weyl
curvature should fall faster than the electric part. Then, the
asymptotic symmetry group reduces to the Poincar\'e group (see, e.g.,
\cite{ar} and references therein) and the asymptotic limits of the
permissible automorphisms in the bulk have to belong to this group.
In this sub-section, we will discuss the analogous restrictions at
$\Delta$.

Recall first that $\Delta$ is foliated by a family of 2-spheres ($v
={\rm const}$) and a transversal direction field $l^{a}$. The
permissible diffeomorphisms are those which preserve this structure.
Hence, on $\Delta$, these diffeomorphisms must be compositions of
translations along the integral curves of $l^{a}$ and general
diffeomorphisms on a 2-sphere in the foliation. Thus, the boundary
conditions reduce ${\rm Diff}\, (\Delta)$ to a semi-direct product of
the Abelian group of ``translations'' generated by vector fields
$\alpha(v) l^{a}$ and ${\rm Diff}\, (S^{2})$. We will refer to this
group as ${\rm Symm}\, (\Delta)$.

The situation with the internal $\SL(2,C)$ rotations is more
subtle. Recall from section \ref{sec2.1} that the 1-form $n$ is tied
to the preferred foliation: given a coordinate $v$ whose level
surfaces correspond to the preferred foliation, we set $n = dv$.
Since the permissible changes in $v$ are of the type $v\mapsto
\tilde{v} = F(v)$, with $F'(v)>0$, the co-vector field $n_{a}$
is now unique up to rescalings $n_{a} \mapsto \tilde{n}_{a}= F'(v)
n_{a}$. Since $(l^a,n_a)$ are normalized via $l^{a}n_{a}\= -1$,
$\Delta$ is now equipped with a class of pairs $(l^a, n_a)$ unique up
to rescalings $(l^a, n_a)\mapsto (\tilde{l^a}, \tilde{n_a}) =
(F'(v))^{-1}l^a, F'(v) {n_a}$).  Hence, given any history $(\sigma,
A)$ satisfying the boundary conditions, we have a spin-dyad $(\i, \o)$
unique up to rescalings
\be \label{gauge}
(\tilde{\i}^{A}, \tilde{\o}^{A}) = ((\exp\Theta)\, \i^{A},
(\exp -\Theta)\, \o^{A}),
\ee
where 
$$\exp\, {\rm Re} (2\Theta) = F'(v) \quad\quad 
{\rm and}\quad \quad\theta := {\rm Im}\, \Theta\quad {\rm
is\,\,\, an\,\,\, arbitrary\,\,\, function\,\,\, on\,\,\,} \Delta .
$$
This suggests that we fix on $\Delta$ a spin-dyad  $(\i, \o)$
{\it up to these rescalings} and allow only those histories in which 
$\sigma^{a}_{AA'}$ maps $(i^{A}\bar{\i}^{A'},\, \o^{A}\bar{\o}^{A'})$
to one of the allowed pairs $(n_{a}, l^{a})$ on $\Delta$. It is easy 
to check that this gauge-fixing can always be achieved. It reduces the
group of local $\SL(2,C)$ gauge transformations to complexified $U(1)$
as in (\ref{gauge}).

As is easy to check, under these restricted internal rotations, the
fields $f(v)$ (which determines the expansion of $n$) and $\alpha,
\beta$ (which determine $\spb{A}$ via (\ref{A})) have the following
gauge behavior:
\be \label{rescaling}
f(v) \mapsto (\exp\, {\rm Re}\,2\Theta)\, f(v) ,\quad
\alpha_{a}\mapsto \alpha_{a}+ \partial_{a}\Theta ,\quad
\beta_{a}\mapsto (\exp\, 2\Theta)\, \beta_{a}\, .
\ee
Thus, $\alpha_{a}$ transforms as a connection while $f$ and $\beta$
transform as ``matter fields'' on which the connection acts. Since $f$
and ${\rm Re}\, \Theta$ are both positive functions of $v$ alone, the
transformation property of $f$ suggests that we further reduce the
gauge freedom to $U(1)$ by gauge-fixing $f$. This is not essential but
it does clarify the structure of the true degrees of freedom and frees
us from keeping track of the awkward fact that ${\rm Re}\, \Theta$
depends only on $v$ while ${\rm Im}\,\Theta$ is an arbitrary
function on $\Delta$ as in local gauge theories. 

Since $f(v)$ has the dimensions of expansion (i.e., $({\rm
length})^{-1}$), and since the only (universally defined) quantity of
the dimension of length is the horizon radius $r_{\Delta}
=(a_{\Delta}/4\pi)^{1/2}$, it is natural to ask that $f$ be
proportional to $1/r_{\Delta}$.  Furthermore, there is a remarkable
fact: for all static black holes in the Einstein-Maxwell theory (with
standard normalization of the Killing field), the expansion of $n$ is
given by $-2/r_{\Delta}$ {\it irrespective} of the values of charges
or of the cosmological constant, so that, in all these solutions,
$f(v) = 1/r_{\Delta}$. This fact can be exploited to extend the
definition of surface gravity to non-static black holes
\cite{abf2}. Although we will not use surface gravity in this paper or
its companion \cite{5}, for uniformity, we will use the same gauge and
set $f(v) = 1/r_{\Delta}$. As is clear from (\ref{rescaling}), this
choice can always be made and furthermore exhausts the gauge freedom
in the real part of $\Theta$. Thus, the internal $\SL(2,C)$ freedom
now reduces to $U(1)$. We wish to emphasize however that none of the
conclusions of this paper or \cite{5} depend on this choice; indeed,
we could have avoided gauge fixing altogether.

Let us summarize. With the gauge fixing we have chosen, only those
automorphisms of the $\SL(2,C)$ spin-bundle in the bulk are permissible
which (reduce to identity at infinity and) belong, on $\Delta$, to the
semi-direct product of the local $U(1)$ gauge group and ${\rm
Sym}\,\Delta$.  Under $U(1)$ gauge rotations, the basic fields
transform as follows:
\be
\alpha_{a} \mapsto \alpha_{a} + i\partial_a\theta, \quad 
\beta_a \mapsto \beta_a, \quad 
\i^{A}\mapsto ( \exp\, i\theta)\, \i^{A}, \quad 
\o^{A}\mapsto (\exp \,-i\theta) \, \o^{A}\, ,
\ee
where $\theta = {\rm Im \Theta}$. We will
see in the next section that these considerations match well with the
action principle which will induce a $U(1)$ Chern-Simons theory on
$\Delta$. As usual, the structure of constraints in the Hamiltonian
theory will tell us which of these automorphisms are to be regarded as
gauge and which are to be regarded as symmetries.

\subsection{Consequences of boundary conditions}
\label{sec3.2}

In this sub-section, we will list those implications of our boundary
conditions which will be needed in the subsequent sections of this
paper and in the companion paper on quantization and entropy
\cite{5}. While some of these results are immediate, others require
long calculations. Derivations are sketched in Appendix \ref{appB}.
For alternate proofs, based on the Newman-Penrose formalism,
see \cite{abf2}.

1. Condition (iv.a) implies that the Lie derivative of the intrinsic
metric on $\Delta$ with respect to $l^{a}$ must vanish; ${\cal
L}_{l}\,\pb{g_{ab}} \= 0$. Thus, as one would intuitively expect, the
intrinsic geometry of isolated horizons is time-independent.  Note,
however, that in general there is no Killing field even in a
neighborhood of $\Delta$. Indeed, since the main conditions (iv)
restrict only on the pull-backs of various fields on $\Delta$, we can
not even show that the Lie derivative of the {\it full} metric
$g_{ab}$ with respect to $l^{a}$ must vanish on $\Delta$; i.e., the
4-metric need not admit a Killing field even on $\Delta$.

Nonetheless, since $l^{a}$ is a Killing field for the {\it intrinsic}
(degenerate) metric $\spb{\,g}$ on $\Delta$, it follows, as already
noted, that the expansion of $l^{a}$ is zero which in turn implies
that the area of the 2-sphere cross-sections S of $\Delta$ is constant
in time. We will denote it by $a_{\Delta}$.

2. Conditions (iv) imply that the pull-back of the four-dimensional 
self-dual connection $\a$ to $\Delta$ has the form%
\footnote{ Note that it is redundant to pull-back forms such as
$\alpha$ and $\beta$ which are defined {\it only} on $\Delta$.  The
derivative operator $\spb{\D}$ is given by: $\spb{{\D}_a}\lambda_A =
\partial_a \lambda_A + A_{aA}{}^B\lambda_B$ where $\partial$ is the
unique derivative flat operator which annihilates $\i$ and $\o$.  Note
that the fields $\i,\o$ and $\spb{A}$ are all defined separately on
the southern and the northern hemispheres of $S$ and related on the
overlap by a local $U(1)$ gauge transformation.}:
\be \label{A}
\spb{A_{a}}^{AB} \= -2\alpha_{a} \i^{(A} \o^{B)} - 
\beta_a \o^{(A}\o^{B)}
\ee
where, as before, $\=$ stands for `` equals at points of $\Delta$ to'',
$\alpha$ is a complex-valued 1-form on $\Delta$,
and the complex 1-form $\beta$ is given by:
\begin{equation}\label{sec3:beta}
\beta_a \= i f(v) \bar{m}_a,
\end{equation}
where $f(v)$ and $\bar{m}_a$ are as in the boundary condition (iv.b).

Let us set 
\be 
\alpha_{a} \=  U_{a} + i V_{a}
\ee
where the one-forms $U$ and $V$ are real. It turns out that $U$ is
completely determined by the area $a_{\Delta}$ of the horizon and
matter fields:
\be\label{mu}
U_a \= \, r_\Delta\,\, \left[\, \frac{2\pi}{a_{\Delta}} 
- \frac{\Lambda}{2} - 4\pi G T_{ab}n^al^b\,\right]\, \spb{n_a}
\ee
The one-form $\beta_a$ is completely determined by the 1-form $V$
and the value $a_{\Delta}$ of area via:
\begin{equation}\label{beta}
D_{[a} \beta_{b]} :\= \partial_{[a} \beta_{b]}
- i V_{[a}\beta_{b]} \= 0 \quad {\rm and}\quad 
{\cal F}_{ab} \= i\beta_{[a}\overline{\beta}_{b]}\, ,
\end{equation}
where $D$ is the covariant derivative operator defined by the $U(1)$
connection $V$ and ${\cal F}$ is its curvature. (As discussed in
section \ref{sec3.3}, $\beta$ transforms as a $U(1)$ matter field,
while $V$ transforms as a $U(1)$ connection so that the first equation
is gauge covariant.)  Thus, the boundary conditions imply that the
pull-back $\spb{A_a}^{AB}$ to $\Delta$ of the four-dimensional
$\SL(2,C)$ connection $\a$ is essentially determined by the real
one-form $V$ and area $a_{o}$ of $\Delta$. Finally, $\sdpb{V}$ has a
simple geometric interpretation: the group of tangent-space rotations
of $S$ is $SO(2)$ and $\sdpb{V}$ is the natural spin-connection on the
corresponding $U(1)$-bundle. More precisely,
\be \label{nu=gamma}
\sdpb{V_a} = -i \sdpb{\Gamma_a}^{AB}\i_A \o_B
\ee
where, as before $\Gamma$ is the spin-connection of the spatial
soldering form $\sigma$ on $M$.

3. Boundary condition (iii) enables us to express the curvature of
$\spb{A_a}^{AB}$ in terms of the pull-back to $\Delta$ of the Riemann
curvature of the four-dimensional $\SL(2,C)$ soldering form $\sigma$
(see \cite{18}, Appendix A):
\be 
\pb{F_{ab}}^{AB} \= -\frac{1}{4}\pb{R_{ab}}\,{}^{cd} \, 
\Sigma_{cd}^{AB}\, .
\ee
It turns out that conditions (iv) and (v) then severely restrict the
Riemann curvature. To spell out these restrictions, it is convenient
to use the Newman-Penrose notation (see Appendix \ref{appB}). First,
the components $\Phi_{00}, \Phi_{01}, \Phi_{10}, \Phi_{02}$ and
$\Phi_{20}$ of the Ricci tensor vanish. Second, the components
$\Psi_0, \Psi_1$ and ${\rm Im}\Psi_2$ also vanish. Third, $\Psi_3$ is
not independent but equals $\Phi_{21}$. Finally, the real part of
$\Psi_2$ is constant on $\Delta$. As a consequence, the following key
equality relating the pull-backs of curvature $F$ and self-dual
2-forms $\Sigma$ holds on $\Delta$:
\ba\label{pbcurv}
\pb{F_{ab}}{}^{AB} \= - {2\pi\over a_{\Delta}}
\pb{\Sigma_{ab}}^{AB} 
-2i (3\Psi_2-2\Phi_{11})\, \spb{n_{[a}} \,\spb{\bar{m}_{b]}}\, \o^A\o^B.
\ea
Note that the first term on the right side is simple and its
coefficient is universal, irrespective of the cosmological constant
and values of electric, magnetic and dilatonic charges or details of
other matter fields present at the horizon.  This fact will play a key
role in this paper as well as \cite{5}. In particular, it will lead us
to a universal action principle. 

The relation (\ref{pbcurv}) tells us that the curvature of the
pulled-back connection $\spb{A}$ is severely restricted. Note however
that the above relation holds {\it only} at points of the isolated
horizon $\Delta$. In the interior of space-time $\M$, curvature can be
quite arbitrary due to the presence of gravitational radiation and
matter fields.  Furthermore, even at points of $\Delta$, the
restriction is only on the pulled-back curvature since the main
boundary conditions refer only to fields defined intrinsically on
$\Delta$. In particular, there is no restriction on the components
$\Psi_{3},\Psi_{4}$ of the Weyl curvature, or on the components
$\Phi_{22},\Phi_{12}$ of Ricci curvature or the scalar curvature even
at points of the boundary $\Delta$. In particular, {\it they need not
be spherically symmetric}.

4. Finally, one can further pull-back (\ref{pbcurv}) to the preferred
2-sphere cross sections of $\Delta$ (i.e., transvect the equation with
$m^{[a} \bar{m}^{b]}$). The result can easily be obtained from
(\ref{pbcurv}) by noting that the second term has zero
pull-back. Thus, 
\be\label{3.5a}
\dpb{F_{ab}}^{AB}
\= -\frac{2\pi}{a_{o}}\, \dpb{\Sigma_{ab}}^{AB}
\ee
This equation will play a key role in specification of the boundary
condition on the phase space variables in section \ref{sec4} and in
the passage to quantum theory in \cite{5}. Finally, the curvature on
the left side of (\ref{3.5a}) is completely determined by the curvature
${\cal F}$ of the $U(1)$ connection $V$:
\be \label{3.5b}
\dpb{F_{ab}}^{AB} \= -2i {\cal F}_{ab}\, \i^{(A}\o^{B)}
\= -4i \partial_{[a} V_{b]} \, \i^{(A}\o^{B)}.
\ee

Let us summarize. As one might expect, boundary conditions on $\Delta$
imply that the space-time fields $(\sigma, A)$ that constitute our
histories are restricted on $\Delta$.  While the restriction is not as
severe as that at the boundary at infinity where $\sigma$ must reduce
to a fixed flat soldering form and $A$ must vanish, they are
nonetheless quite strong. Given the constant $a_{o}$ ---the value of
the horizon area--- the only unconstrained part of $\spb{A}$ is the
1-form $V$ and the pull-back $\sdpb{\Sigma}$ of $\Sigma_{ab}^{AB}$
to the 2-sphere cross-sections is completely determined by the
curvature of $V$ by equations (\ref{3.5a}, \ref{3.5b}).

\section{Action and phase space} 
\label{sec4}

Recall from section \ref{sec2} that in the absence of internal
boundaries the action of general relativity is given by:
\be
\label{action1}
S'_{\rm Grav}[\sigma,A]=
-{i\over 8\pi G}\,\,\big[\int_{\cal M}\,{\rm Tr}\left(\Sigma
\wedge F\right) - \int_{\cal T} {\rm Tr}\left( \Sigma\wedge A \right)
\big]  
\ee
Therefore, one might imagine that the presence of the internal
boundary $\Delta$ could be accommodated simply by replacing the
time-like cylinder ${\cal T}$ at infinity in (\ref{action1}) by ${\cal
T}\cup \Delta$.  However, this simple strategy does not work; that
action fails to be functionally differentiable at $\Delta$ because the
boundary conditions at $\Delta$ are quite different from those on
${\cal T}$. In section \ref{sec4.1} we will show that the action can
in fact be made differentiable by adding to it a Chern-Simons term at
$\Delta$. In section \ref{sec4.2} we will perform a Legendre
transform, obtain the phase space and analyze the notion of gauge in
the Hamiltonian framework. Finally, in section \ref{sec4.3}, we will
perform a canonical transformation to obtain a Hamiltonian description
(along the lines of section \ref{sec2.2}) in which all fields are
real.

Since our primary motivation is to construct a Hamiltonian framework
which will serve as a point of departure for the entropy calculation
in the next paper \cite{5}, we will confine ourselves to histories
with a {\it fixed} value of isolated horizon parameters
\footnote{A treatment which allows fields with arbitrary values of
parameters is necessary in order to generalize the laws of black hole
mechanics and is given in \cite{abf2}.}.
In this section, the only parameter is the area $a_\Delta$ (or, the
radius $r_\Delta$, where $a_\Delta = 4\pi r_\Delta^2$). In the next
section, we will also fix the electric, magnetic and dilatonic
charges.  In the classical theory we will thus be led to a phase
space, each point of which admits an isolated horizon with given
values of parameters.  The idea is to quantize this sector in a way
that allows for appropriate quantum fluctuations also at the boundary
\cite{5}. The surface states at the horizon in the resulting quantum
theory will account for the entropy of a black hole (or cosmological
horizon) with the specified horizon area and charges.

\subsection{Action Principle}
\label{sec4.1}

Consider a 4-manifold ${\cal M}$, topologically $M\times R$. We will
work with a fixed cosmological constant $\Lambda$, i.e. with a fixed
theory. If $\Lambda \le 0$, the `spatial' 3-manifold $M$ will be taken
to be diffeomorphic to $S^2\times R$ with an internal boundary $S$
with a 2-sphere topology, while if $\Lambda >0$, it will be taken to
be the complement of an open ball in $S^3$, again with an internal
boundary $S$ with a 2-sphere topology. Consider on ${\cal M}$ smooth
histories $(A,\sigma)$ satisfying suitable boundary conditions. 
There are conditions at infinity which require the
fields to be asymptotically flat in the standard sense \cite{ar}  if
$\Lambda =0$, and asymptotically anti-de Sitter if $\Lambda <0$
\cite{am}.  In all cases, the boundary conditions at the internal
boundary $\Delta =S\times R$ will be the isolated horizon conditions
spelled out in Section \ref{sec3.1}. While all main considerations go
through irrespective of the value of the cosmological constant,
as in Section \ref{sec2}, for definiteness we will set $\Lambda =0$
in the main discussion.%
\footnote{In the final description, the case with $\Lambda
\not=0$ can be recovered by adding the obvious term ($ \Lambda {\rm
Tr} \sigma \wedge\sigma \wedge \sigma$) to the scalar (or Hamiltonian)
constraint, and, ignoring the terms at infinity if $\Lambda >0$ and
replacing them appropriately \cite{am} if $\Lambda <0$.}%

Let us begin with the action $S'_{\rm Grav}(A,\sigma)$. Fix a region of ${\cal
M}$ bounded by two (partial) Cauchy surfaces $M_1$ and $M_2$ which
extend to spatial infinity in the asymptotic region and intersect the
isolated horizon $\Delta$ in two 2-spheres $S_1$ and $S_2$ of our
preferred foliation. Since $\sigma$ appears undifferentiated in the
action, and since we can replace $\Sigma$ in (\ref{action1}) by its
fixed boundary value on $\cal T$ in the surface term, the variation with respect
to $\sigma$ is well-defined and gives rise only to the bulk equation
of motion $\sigma^a_{AA'}F_{ab}^{AB} = 0$. The variation with respect
to $A$, on the other hand, gives rise to a surface term
\be \label{var1}
[\delta S'_{\rm Grav}]_\Delta = - {i\over 8\pi G}\, \int_\Delta
{\rm Tr}\,\, \Sigma \wedge \delta A \ee
Now, the boundary condition (\ref{A}) implies that, for every $A$ in
our space of histories, $A_a^{AB} \o_A \o_B \=0$. Hence, $\delta A_a^{AB}
\o_A\o_B \=0$. We can now use (\ref{pbcurv}) to conclude that
\be 
{\rm Tr}\,\, \spb{\Sigma}\wedge \spb{\delta A} = - {a_\Delta\over 2\pi}
\,\,\,{\rm Tr} \,\,\spb{F}\wedge \spb{\delta A}.
\ee
Since the 3-form in the integrand of (\ref{var1}) is pulled back to
$\Delta$, we have:
\ba \label{var2}
[\delta S'_{\rm Grav}]_\Delta  &=& {i\over 8\pi G}\,\,
{a_\Delta\over 2\pi} \int_\Delta {\rm Tr}\,\, F \wedge \delta A 
\nonumber \\
&=& {i\over 8\pi G}\,\, {a_\Delta\over 4\pi}
\delta \int_\Delta {\rm Tr}\,(A\wedge dA + {2\over 3} A\wedge A 
\wedge A)\, , \nonumber \\
\ea
where in the second step we have used the fact that, since $\delta A$
vanishes on $M_1$ and $M_2$, it vanishes on the 2-spheres $S_1$ and
$S_2$ on $\Delta$.  Note that the right side is precisely the
variation of the Chern-Simons action for the connection $\spb{A}$ on
$\Delta$. Hence, it immediately follows that the action
\ba \label{S}
S_{\rm Grav} (A,\sigma) &:=& S'_{\rm Grav}(A,\sigma) - {i\over 8\pi G}
\,\, {a_\Delta \over 4\pi}\,\,
S_\Delta^{\rm CS}\nonumber \\
&=&  - {i\over 8\pi G}\left[ \int_{\cal M} {\rm Tr}\, \Sigma\wedge F
- \int_{\tau}{\rm Tr}\, \Sigma\wedge A + {a_\Delta\over 4\pi}
\int_\Delta {\rm Tr}\, (A\wedge dA + {2\over 3} A \wedge A\wedge A)\right]
\nonumber \\
\ea
has a well-defined variation with respect to $A$ which gives rise only
to the bulk equation of motion $\D \wedge \Sigma =0$. 
 
To summarize, with our boundary conditions at infinity and at the
isolated horizon $\Delta$, the action $S_{\rm Grav}(\sigma, A)$ of
(\ref{S}) is differentiable and its variation yields {\it precisely}
Einstein's equations on ${\cal M}$. In spite of the presence of
boundary terms, there are no additional equations of motion either at
the time-like cylinder $\tau$ at spatial infinity or on the isolated
horizon $\Delta$. In particular, although the boundary term at
$\Delta$ is the Chern-Simons action for $\spb{A}$, we do {\it not}
have an equation of motion which says that the curvature $\spb{F}$ of
$\spb{A}$ vanishes.  Indeed, $\spb{F}$ is no-where vanishing and is
given by (\ref{pbcurv}).  Nonetheless, we will see that the presence
of this boundary term does give rise to an addition of the
Chern-Simons term to the symplectic structure of the theory, which in
turn plays a crucial role in the quantization procedure. Thus, the
role of the surface term $S^{\rm CS}_\Delta$ is subtle but important.

{\it Remarks:}

a) The boundary conditions in section \ref{sec3} were motivated by
geometric considerations within general relativity and capture the
idea that $\Delta$ is a non-rotating, isolated horizon. The fact that
these then led to a consistent action principle is quite non-trivial
by itself. The fact that the added boundary term has a simple
interpretation as the Chern-Simons action for the self-dual connection
$\spb{A}$ on $\Delta$ is remarkable. We will see that this delicate
interplay between classical general relativity and Chern-Simons theory
also extends to quantum theory, where the matching extends even to
precise numerical factors.

b) Note that the Chern-Simons action arose because of equations
(\ref{A}) and (\ref{pbcurv}). These equations have a ``universality'':
the inclusion of a cosmological constant or electric, magnetic and
dilatonic charges have no effect on them. Consequently, in all these
cases, the coefficient of the Chern-Simons term is always
$a_\Delta/4\pi$. It is likely that this ``universality'' is directly
related to the ``universality'' of the expression $S_{\rm bh} =
a_\Delta/4\ell_P^2$ of the Bekenstein-Hawking black hole entropy in
general relativity. 

c) We can now see in detail why we could not have simply replaced
$\tau$, the time-like cylinder at infinity, by ${\cal \partial M}=
\tau \cup \Delta$ in the expression (\ref{action1}) of $S'_{\rm Grav}$
to obtain a well-defined action principle. At infinity, the soldering
form $\sigma$, and hence the 2-forms $\Sigma$, are required to
approach their values in the background flat space as $1/r$ and the
connection $A$ falls off as $1/r^2$. Hence the variation of the
surface term in $S'_{\rm Grav}$ with respect to $\sigma$ vanishes
identically on $\tau$. On the inner boundary $\Delta$, by contrast,
the 2-forms $\spb\Sigma$ are {\it not} fixed. Instead, their values
are tied to those of $\spb{F}$. Hence the simple replacement of $\tau$
by $\tau \cup \Delta$ does not yield a differentiable action.  Indeed,
we may re-express $S_\Delta^{CS}$ using $\spb\Sigma$ and
$\spb{A}$. The result is 
${1\over 2} \int_\Delta {\rm Tr}\,(\Sigma\wedge A)$ 
rather than $\int_\Delta {\rm Tr}\,(\Sigma\wedge A)$. Thus, while the
boundary terms at $\tau$ and $\Delta$ can be cast in the same form,
they differ by a factor of 2. Therefore, contrary to what is sometimes
assumed, the total action {\it can not} be expressed as a volume
integral, i.e. one can not get rid of all surface terms using Stokes'
theorem. This seems surprising at first.  However, such a situation
arises already in the case of a scalar field in flat space if there
are two boundaries and one imposes the Dirichlet conditions at one
boundary and the Neumann conditions at the other.

d) There are several contexts in which a Chern-Simons action has
emerged as a boundary term. In some of these discussions, one begins
with a theory, notices that the action is not fully gauge invariant
and adds new boundary degrees of freedom to obtain a more satisfactory
action for the extended system. The boundary degrees are typically
connections and their dynamics is governed by the Chern-Simons action
(see, in particular, \cite{12,13}).  By contrast, in the present work
we did not add new degrees of freedom at all. The Chern-Simons piece
did arise because the naive action $S'_{\rm Grav}(A,\sigma)$ fails to
admit a well-defined variational principle. However, the effect of the
boundary conditions is to {\it reduce} the number of degrees of
freedom: as usual, the boundary conditions impose relations between
dynamical variables which are independent in the bulk. Furthermore,
unlike in other discussions, these conditions arose from detailed
geometrical properties of null vector fields $l$ and $n$ associated
with isolated horizons in general relativity. This is also a major
difference from the discussion in \cite{ls} where the Chern Simons
action arose as a boundary term in {\it Euclidean} gravity subject to
the condition that the spin-connection reduce to a right-handed
$\SU(2)$ connection on the boundary. Also, while in other contexts the
boundary connections are non-Abelian (typically $\SU(2)$), as noted in
Section \ref{sec3}, in our case, the independent degrees of freedom on
$\Delta$ are coded in the Abelian connection $V$. This fact will play
an important role in quantization.
\bigskip

We will conclude this section by expressing the action in terms of
this Abelian connection. Using the expression (\ref{A}) of the
connection $\spb{A}$ on the boundary $\Delta$ and the expression
(\ref{mu}) of $U$, it is easy to verify that the
action can be re-written as
$$  S_{\rm Grav} (A,\sigma) = 
S'_{\rm Grav}(A,\sigma) + {i\over 8\pi G}\, {a_\Delta\over
2\pi}\, \int_\Delta V\wedge dV \,\,- \,\, {1\over 8\pi G}
\int_\Delta U \wedge\,{}^{2}\epsilon $$
where ${}^2\epsilon$ is the area 2-form on the preferred 2-spheres.
It is easy to verify that, since $a_\Delta$ is fixed on the space of
paths, the variation of the last integral vanishes in the
Einstein-Maxwell-Dilaton theory. Hence,
$$ \label{Abel} \tilde{S}_{\rm Grav} (A,\sigma) = 
S'_{\rm Grav}(A,\sigma) + {i\over 8\pi G}\, {a_\Delta\over 2\pi}\, 
\int_\Delta V\wedge dV . $$
is also a permissible action which yields the same equations of motion
as $S_{\rm Grav}$. (In passing from $S_{\rm Grav}$ to $\tilde{S}_{\rm
Grav}$ we have merely used the freedom to add to the action a function
of dynamical variables which is constant on the space of paths.)  In
this form, we see explicitly that the surface term at $\Delta$ is the
Chern-Simons action of the {\it Abelian} connection $V$. It is not
surprising that the action depends only on the Abelian part $V$ of the
$\SL(2,C)$ connection $\spb{A}$ since $\spb{A}$ is completely
determined by $V$ on our space of paths.  However, it is pleasing that
the functional dependence of the action on $V$ is simple, being just
the Chern-Simons action for $V$.

Finally, the Abelian nature of $V$ implies that, like other terms
in the action, the Chern-Simons piece is also fully gauge invariant;
the usual problem with large gauge transformations does not arise.

\subsection{Legendre transform, constraints and gauge}
\label{sec4.2}

We are now ready to pass to the Hamiltonian framework by performing
the Legendre transform. Consider any history $A, \sigma$ on ${\cal M}$
and introduce a time function $t$ on ${\cal M}$ such that: i) each
leaf $M(t)$ of the resulting foliation is space-like; ii) at infinity
$t$ reduces to a Minkowskian time coordinate of the flat fiducial
metric near $i^o$, each $M(t)$ passes through $i^o$ and the pull-backs
of the dynamical fields $A, \Sigma$ to $M(t)$ are asymptotically flat,
i.e., satisfy (\ref{bc1}); and, iii) on the isolated horizon $\Delta$,
the time-function $t$ coincides with the function $v$ labeling the
preferred 2-spheres $S(t)$, and the unit normal $\tau^a$ to $M(t)$ is
given by $\tau^a \= (l^a+n^a)/\sqrt{2}$ on each $S(t)$. Next,
introduce a `time-evolution' vector field $t^a$ such that: i) ${\cal
L}_t t = 1$; ii) at infinity, $t^a$ is orthogonal to the leaves
$M(t)$, i.e., $t^a = N\tau^a$ at $i^o$ for some `lapse function' $N$;
and, iii) on $\Delta$, $t^a \= l^a$. The conditions on $t^a$ at the
two boundaries imply that $t$ and $t^a$ define the same asymptotic
rest frame at infinity and, at $\Delta$, the frame coincides with the
`rest frame' of the isolated horizon. These two restrictions are not
essential; they are introduced just to avoid some minor technical
complications which are inessential to our main discussion.

Since the action is written in terms of forms, in contrast to 
the standard calculation in geometrodynamics, the Legendre transform
is almost trivial to perform. One obtains:
\ba \label{lt} 8\pi i G \,S_{\rm Grav}(A,\sigma) = &-&\int
dt\int_{M(t)} {\rm Tr}\,\, \left(\Sigma \wedge {\cal L}_t A + (A.t) \D
\Sigma - \Sigma \wedge (\vec{N}\cdot F) + iN\sqrt{2} \sigma \wedge
F\right) \nonumber\\ &-& \int dt\oint_{{S_\infty}(t)} {\rm Tr}\,\,
i\sqrt{2}N\, \sigma\wedge A +{a_\Delta\over 4\pi}\int dt
\oint_{{S_\Delta}} {\rm Tr}\,\, A\wedge {\cal L}_l A\nonumber \\
\ea
where, as usual, the lapse $N$ and the shift ${\vec N}$ are defined
via $t^a = N \tau^a +{\vec N}^a$, the 1-form $\vec{N}\cdot F$ is
defined via $(\vec{N} \cdot F)_b := {\vec N}^a F_{ab}$, and $\sigma$ is
the spatial, $SU(2)$ soldering form on $M(t)$ as in Section
\ref{sec2}. (In terms of $\Sigma$, we have $\sqrt{2}i \sigma_m =
\tau^a\,\Sigma_{ab}q_m{}^b$ where $q_m{}^b$ is the projection
operator on $M(t)$.)  The surface term can be re-expressed in terms of
the $U(1)$ connection $V$ as:
\be {a_\Delta\over 4\pi}\int dt \oint_{{S_\Delta}(t)} {\rm Tr}\,\,
A\wedge {\cal L}_l A = - {a_\Delta \over 2\pi} \int dt
\oint_{{S_\Delta}} V \wedge {\cal L}_l V \ee

From the Legendre transform (\ref{lt}) it is straightforward to obtain
the phase-space description. Denote by $M$ a generic leaf of the
foliation. It is obvious that the dynamical fields are the pull-backs
to $M$ of pairs $(A, \Sigma)$.%
\footnote{From now on, in this section we will denote these
pulled-back fields simply by $A$ and $\Sigma$. The 4-dimensional
fields on ${\cal M}$ will carry a superscript 4 (e.g. ${}^4\!A$) to
distinguish them from the 3-dimensional fields on $M$.}
Note that there are no independent, surface degrees of freedom either
at infinity or on the horizon: since all fields under consideration
are smooth, by continuity, their values in the bulk determine their
values on the boundary.%
\footnote{By contrast, in quantum theory, the relevant histories are
distributional and hence values of fields in the bulk do not determine
their values on the boundary. We will see in \cite{5} that it is this
fact leads to quantum surface states which in turn account for the
black hole entropy.}
In fact, as usual, the boundary conditions serve to {\it reduce} the
number of independent fields on $S_\infty$ and $S_\Delta$. At
infinity, the fields $A, \Sigma$ on $M$ must satisfy the asymptotic
conditions (\ref{bc1}); in particular, their limiting values are
totally fixed. On the horizon, the pull-back of the connection to
$S_\Delta$ is of a restricted form dictated by (\ref{A}):
\be \sdpb{A_a}{}^{AB} \= -2iV_a \i^{(A} \o^{B)} + \beta_a \o^A\o^B
\ee
where $\beta$ is determined by $V$ via (\ref{mu}). Furthermore, the
curvature ${\cal F}$ of $V$ is related to $\sdpb{\Sigma}$ via:
\be \label{calF}
{\cal F} \= {2\pi i \over a_\Delta}\,\, \sdpb{\Sigma}^{AB} 
\i_A\o_B
\ee
Since $\beta = (i/r_\Delta) \overline{m}$, these two equations
together with (\ref{mu}) imply that $V$ is the only independent
dynamical field on $S_\Delta$.

The symplectic 1-form $\Theta$ is easily obtained from the Legendre
transform. We have:
\be \label{Theta}
8\pi i G \,\,\, \Theta|_{(A,\sigma)}(\delta) = 
-\int_M {\rm Tr}\,\delta A \wedge \Sigma \,+\,\, 
{a_\Delta \over 4\pi} \oint_S {\rm Tr}\,\delta A \wedge A\,. 
\ee
The symplectic structure $\Omega$ is just the exterior derivative of
$\Theta$:
\ba 
8\pi iG\,\, \,\Omega|_{(A,\sigma)} (\delta_1, \delta_2) &=&
\int_M {\rm Tr}\,\, (\delta_1 A \wedge \delta_2 \Sigma
-\delta_2 A \wedge \delta_1 \Sigma )
- {a_\Delta\over 2\pi} \oint_{S} {\rm Tr}\,\delta_1 A\wedge 
\delta_2 A \nonumber\\
&=&\int_M {\rm Tr}\,\, (\delta_1 A \wedge \delta_2 \Sigma
-\delta_2 A \wedge \delta_1 \Sigma)
+ {a_\Delta\over \pi} \oint_{S} \delta_1 V\wedge \delta_2 V
\nonumber\\
\ea
for any two tangent vectors $\delta_1 \equiv (\delta_1 A,
\delta_1\Sigma)$ and $\delta_2 \equiv (\delta_2 A, \delta_2\Sigma)$ at
$(A,\Sigma)$ to the phase space. Finally, it is clear from the
Legendre transform (\ref{lt}) that the fields $({}^4\!A.t), {\vec N},
N$ are Lagrange multipliers. The resulting constraints are the
standard ones:
\be
\D_a \tilde\sigma^a = 0, \quad \quad  
{\rm Tr}\, \tilde\sigma^a F_{ab}= 0,  \quad\quad {\rm and} 
\quad \quad {\rm Tr}\tilde\sigma^a\tilde\sigma^bF_{ab}=0\, ,
\ee
where $\tilde\sigma^a$ is essentially the dual of the 2-form
$\Sigma_{ab}$ on $M$: $2 \sqrt{2} \tilde\sigma^a =
\tilde\eta^{abc}\Sigma_{bc}$ where $\tilde{\eta}$ is the metric
independent Levi-Civita density on $M$. As usual, these form a 
set of first class constraints.

Recall that, in the Hamiltonian description, first class constraints
generate gauge.  Let us therefore examine the constraints one by one.
Smearing the first (Gauss) constraint by a field $\lambda_A{}^B$ and
integrating over $M$, we obtain a function on the phase space:
\be 8\pi iG\,\, {\cal C}_\lambda (A,\Sigma) := 
\int_M {\rm Tr}\,\, \lambda \D\Sigma
\ee
whose variation along a general vector $\delta$ at a point $(A,
\Sigma)$ yields
\be 8\pi i G \,\, \delta {\cal C}_\lambda = \int_M {\rm Tr}\left(
\lambda\, \delta A \wedge \Sigma - \lambda \Sigma\wedge \delta A -
\D\lambda \wedge \delta\Sigma\right) 
+ \oint_{\partial M} {\rm Tr}\,\lambda \delta\Sigma\, .  \ee
The question is if $\delta{\cal C}_\lambda$ is of the form
$\Omega(\delta, \delta_\lambda)$ for some tangent vector
$\delta_\lambda$. If so, $\delta_\lambda$ would be the Hamiltonian
vector field generated by the constraint functional $C_\lambda$.
Alternatively, since $\delta {\cal C}_\lambda$ vanishes for all vectors
$\delta$ tangential to the constraint surface, $\delta_\lambda$ will
be a degenerate direction of the pull-back of $\Omega$ to the
constraint surface. Each of these properties implies that
$\delta_\lambda$ would represent an infinitesimal gauge motion in the
Hamiltonian theory.

A short calculation yields
\be \label{gt1}
\delta {\cal C}_\lambda = \Omega(\delta, \delta_\lambda) 
\ee
for all tangent vectors $\delta$, with 
\be \label{gt2}
\delta_\lambda  = (\D\lambda,\,\, [\Sigma, \lambda])\, ,
\ee
{\it provided}: i)$\lambda_A{}^B$ tends to zero at infinity (i.e.  is
$O(1/r)$), and, ii) has the form $\lambda (\i^A\o_B + \o^A\i_B)$ on
$\Delta$.  Note that the two conditions are necessary to ensure that
$\delta_\lambda$ is a well-defined tangent vector to the phase
space. Furthermore, (\ref{gt2}) is precisely an internal $SL(2,C)$
rotation compatible with our boundary conditions. As one would have
expected, the Hamiltonian theory tells us that these should be
regarded as gauge transformations of the theory. Technically, the only
non-trivial point is that, {\it because of the presence of the surface
term in the symplectic structure, we do not have to require that
$\lambda_A{}^B$ should vanish on $\Delta$ for (\ref{gt1}) to
hold}. The Gauss constraint generates internal rotations which can be
non-trivial on the horizon.

Next, let us consider the `diffeomorphism constraint'. The analysis
\cite{18} in the case without boundaries suggests that we consider
the constraint function $C_{\vec{N}}$ defined by
\be
8\pi iG\,\, {\cal C}_{\vec N}(A,\Sigma) := -\int_{M} {\rm Tr}\, 
\left( \Sigma\wedge\vec{N}.F - (A.\vec{N}) \D\Sigma \right)
\ee
where the smearing field $\vec{N}$ is a suitable vector field on $M$.
The variation of this function along an arbitrary tangent vector 
$\delta$ to the phase space at the point $(A,\Sigma)$ yields,
\be 8\pi i G \,\,\delta {\cal C}_{\vec{N}} = \int_M {\rm Tr}\,\left(
\delta A \wedge {\cal L}_{\vec{N}} \Sigma - {\cal L}_{\vec{N}} A
\wedge \delta\Sigma\right) - \oint_S {\rm Tr}
\left(\delta(A.\vec{N})\, \Sigma + A.\vec{N} \delta{\Sigma}\right)\, .
\ee
It is easy to verify that
\be 
\delta{\cal C}_{\vec N} =  \Omega(\delta, \delta_{\vec N})
\ee
for all $\delta$, with
\be 
\delta_{\vec N} = ({\cal L}_{\vec{N}} A,\,\, {\cal L}_{\vec{N}} \Sigma)\, ,
\ee
provided the smearing field ${\vec N}$ satisfies the following
properties: i) it vanishes (as $O(1/r)$) at infinity; and, ii) it is
tangential to $S$. Again, note that the smearing field ${\vec N}$ does 
{\it not} have to vanish on $S$; it only has to be tangential to $S$. 
The diffeomorphisms generated by all such vector fields 
$\vec{N}$ are to be regarded as `gauge' transformations in this
Hamiltonian theory. Note that asymptotic translations or rotations
have well-defined action on the phase space but they are not generated
by constraints and therefore are not regarded as gauge. This is the
standard situation in the asymptotically flat context. On the internal
boundary, diffeomorphisms which fail to be tangential to the boundary
also do not correspond to gauge. This is not surprising since such
diffeomorphisms do not even give rise to well-defined motions on the
phase space.

Finally let us consider the scalar (or the Hamiltonian) constraint
smeared by a lapse function $N$:
\be 8\pi iG \,\,{\cal C}_N :=  i\sqrt{2}\int_M {\rm Tr}\, 
N\sigma\wedge F 
\ee
The analysis is completely parallel to that of the other two constraints.
The result is:
\be 
\delta {\cal C}_N  = \Omega (\delta, \delta_N) 
\ee
for all $\delta$, with
\be \label{hameq}
\delta_N = ({N\over 4}\epsilon^{bmn} 
[\Sigma_{mn}, F_{ab}]\,\, , \D_{[b} N\sigma_{c]})\, , 
\ee
provided the lapse $N$ tends to zero both at infinity and on $S$.
(Here $\epsilon^{abc}$ is the 3-form defined by the spatial soldering
form $\sigma$ and the square bracket denotes the commutator with
respect to spinor indices.) In this case, (\ref{hameq}) are precisely
the Hamilton's equations of motion for the basic canonical variables.

To summarize, the smearing fields $\lambda, {\vec N}, N$ have to
satisfy certain boundary conditions for the corresponding constraint
functions to generate well-defined canonical transformations.
Therefore, as in the case without internal boundaries \cite{18}, the
constraint sub-manifold of the phase space is defined by the vanishing
of the constraint functions ${\cal C}_\lambda, {\cal C}_{\vec N},
{\cal C}_N$ where the smearing fields satisfy these conditions.  The
corresponding canonical transformations represent gauge motions.  In
quantum theory, physical states are to be singled out by requiring
that they be annihilated by the quantum operators corresponding to
{\it these} constraint functions.

\subsection{Passage to real variables}
 \label{sec4.3}

As explained in section \ref{sec2.2}, at the present stage of
development, quantization is fully manageable only in terms of real,
$\SU(2)$ connections. Therefore, as in section \ref{sec2.2} we will
now carry out the canonical transformation to manifestly real
variables, paying attention to the internal boundary $\Delta$ and
taking in to account the presence of the boundary term in the
symplectic structure.

Recall first that we can always express the connection $A$ on $M$ as
$A_a = \Gamma_a - (i/\sqrt{2})K_a$ for some $K_a^{AB}$, where $\Gamma_a^{AB}$
is the spin-connection compatible with the spatial soldering form
$\sigma$.  Furthermore, the fact that we are working with the real,
Lorentzian theory implies that $\Gamma$ is a real $\SU(2)$ connection
and $K_a$ is a real, $\su(2)$-valued 1-form.  (When equations of
motion are satisfied, $K_{ab} := - {\rm Tr}K_a \sigma_b$ has the
interpretation of the extrinsic curvature of $M$.) Therefore, the
symplectic 1-form $\Theta$ of (\ref{Theta}) can be written as:
\ba \label{1f} 
8\pi i G\,\, \Theta(\delta) &=& - \int_M {\rm Tr}\,\delta
A\wedge \Sigma \,+\, 
{a_\Delta\over 4\pi}\oint_S {\rm Tr}\,\delta A\wedge
A\nonumber\\ 
&=& ({i\over \sqrt{2}}) \int_M {\rm Tr}\,\delta K \wedge
\Sigma - \int_M {\rm Tr}\,\delta\Gamma \wedge\Sigma - 
{a_\Delta\over 2\pi} \oint_S  \delta {V}\wedge V \nonumber\\
\ea

Now, it is well-known \cite{18} that the integrand of the second term
is an exterior derivative of a 2-form which vanishes at infinity.
Therefore, that term can be written \cite{29} as an integral over the inner
boundary $S$:
\be \label{4.27}
\int_M {\rm Tr}\, \delta\Gamma \wedge\Sigma = -{1\over 2}
\oint_S {\rm Tr}\, \sigma \wedge \delta\sigma
\ee
Thus, $\Theta = \Theta^{(M)} + \Theta^{(S)}$, where
\ba \label{4.28} 
8\pi G \Theta^{(M)}\,\,(\delta) &=& {1\over \sqrt{2}}
\int_M {\rm Tr}\, \delta K \wedge \Sigma,\nonumber\\ 
8 \pi i G \Theta^{(S)} (\delta) &=&  {1\over 2}
\oint_S {\rm Tr}\, \sigma \wedge \delta\sigma
+ {a_{\Delta}\over 4\pi} \oint {\rm Tr}\, \delta A 
\wedge A\nonumber\\
\ea
Using this fact, we will now show that, as one would expect, the
symplectic structure is real. 

For this, we first note that our boundary conditions imply that any
tangent vector $\delta$ to our phase space has the following form when
evaluated on $S$:
\be \label{4.29}
(\delta \sdpb{A},\,\, \delta \sdpb{\sigma}) 
\= \left( [{\sdpb{\sigma}}\,, \lambda] +{\cal L}_\xi\, \sdpb{\sigma}, 
\,\,\,
\sdpb{\D}\lambda + {\cal L}_\xi\,\sdpb{A}\right)
\ee
where $\lambda^{AB} = 2ih \i^{(A} \o^{B)}$ for some real function $h$
on $S$ and $\xi$ is a vector field on $S$ satisfying $\oint 
{\cal L}_\xi\,\,{}^2\epsilon \, =0$. Next, recall that, since 
the symplectic structure $\Omega$ is the exterior derivative 
of $\Theta$, we have
\be\label{4.30}
\Omega(\delta_1, \delta_2) = {\cal L}_{\delta_1}\,\,\Theta(\delta_2)
- {\cal L}_{\delta_2}\,\, \Theta(\delta_1) - \Theta 
([\delta_1, \delta_2])
\ee
for any two vector fields $\delta_1, \delta_2$ on the phase space.
Using (\ref{4.29}), one can now show that $\Omega^{(S)}$, obtained by
substituting $\Theta^{(S)}$ for $\Theta$ in (\ref{4.30}),
vanishes. Hence $\Theta^{(S)}$ is closed and does not contribute to
the symplectic structure. Since $\Theta^{(M)}$ is manifestly real, so
is the symplectic structure.

Finally, as in Section \ref{sec2.2}, let us define real,
$\SU(2)$-valued forms
\begin{eqnarray}
\Ag_a{}^{AB} := \Gamma_a{}^{AB} - {\gamma\over \sqrt{2}} K_a{}^{AB} 
\nonumber \\
\Eg_{ab}{}^{AB} := {1\over {\gamma}} \, \Sigma_{ab}{}^{AB}.
\end{eqnarray}
Since $\Theta^{(S)}$ is curl-free, it follows that the one form
\ba 
8\pi  G \, \, {}^\gamma\!\Theta(\delta) &=& {\gamma\over \sqrt{2}}
\int_M {\rm Tr}\, \delta K \wedge \Eg - \int_M {\rm Tr}\, 
\delta\Gamma\wedge\Eg + {a_\Delta\over 4\pi\gamma} 
\oint_S {\rm Tr}\, \delta {\Ag}\wedge \Ag \nonumber\\
&=& - \int_M {\rm Tr}\, \delta \Ag\wedge \Eg + 
{a_\Delta\over 4\pi\gamma}\oint_S {\rm Tr}\, \delta \Ag\wedge
\Ag \nonumber\\ 
\ea
is also a symplectic potential for the total symplectic structure 
$\Omega$. Hence we have:
\ba \label{Omega}
8\pi G \, \Omega|_{(\Ag,\Eg)} (\delta_1, \delta_2) &=&
\int_M {\rm Tr}\,\, (\delta_1 \Ag \wedge \delta_2 \Eg
-\delta_2 \Ag \wedge \delta_1 \Eg )
- {a_\Delta\over 2\pi\gamma} \oint_{S} \delta_1 \Ag\wedge 
\delta_2 \Ag \nonumber\\
&=&\int_M {\rm Tr}\,\, (\delta_1 \Ag \wedge \delta_2 \Eg
-\delta_2 \Ag \wedge \delta_1 \Eg )
+ {a_\Delta\over \pi\gamma} \oint_{S} \delta_1 V\wedge \delta_2 V
\nonumber\\
\ea

Thus, as in section \ref{sec2.2}, we now have a phase space
formulation in terms of manifestly real variables. The phase space
variables $(\Ag, \Eg)$ are subject to boundary condition (\ref{bc2})
at infinity. On the horizon boundary $S$, the independent part of
$\Ag$ is contained in a ($\gamma$-independent) $\U(1)$ connection $V$,
whose curvature completely determines the pull-back to the boundary
$S$ via (\ref{3.5a}). The symplectic structure is given by
(\ref{Omega}). As mentioned above, this formulation in terms real
variables is not necessary for any of the classical considerations
---such as the laws of isolated horizon mechanics--- but is needed, at
this stage, for the passage to quantum theory \cite{5}.

\section{Inclusion of hair: electric, magnetic and dilatonic 
charges}
\label{sec5}

The boundary conditions at the horizon specified in section \ref{sec3}
allow a non-zero cosmological constant and matter fields are subject
only to condition (v) in the main Definition. Once this condition is
satisfied, the discussion of section \ref{sec4} goes through and is
completely insensitive to the details of matter fields.  However, in a
complete theory, we also need to ensure that the matter action is
differentiable and work out the Hamiltonian framework for the matter
sector as well. This can require imposition of additional boundary
conditions on matter fields.  In this section, we will incorporate
Maxwell and dilatonic fields.  The Einstein-Maxwell case is discussed
in section \ref{sec5.1} and the dilatonic case in section
\ref{sec5.2}.

\subsection{Maxwell Fields}
\label{sec5.1}

Since we require that field equations hold on $\Delta$, the
gravitational boundary conditions already imply certain restrictions
on the behavior of Maxwell fields there.  Let us begin with these.  As
noted in section \ref{sec3.2}, boundary conditions imply that several
components of the Ricci tensor vanish on $\Delta$. In particular,
since the expansion of $l^a$ vanishes, the Raychaudhuri equation
implies that the Ricci tensor must satisfy $R_{ab}l^a l^b \=0$. Since
$R_{ab} - \frac{1}{2}R g_{ab} +\Lambda \, g_{ab} \= 8\pi G\, T_{ab}$,
the stress-energy tensor of the Maxwell field must satisfy $T_{ab}l^a
l^b \= 0$ in the Einstein-Maxwell theory.

To find implications of this condition for the electro-magnetic field
${\bf F}$, it is convenient to first recast it in the spinorial
form. ${\bf F}$ can be decomposed over the basis of 2-forms
$\Sigma_{ab}, \overline{\Sigma}_{ab}$ (see Eq (\ref{sigma}) in
Appendix A) as:
\begin{equation}\label{ch:1}
-2 {\bf F}_{ab} = {\bf \phi}_{AB} \Sigma_{ab}^{AB} + 
\overline{{\bf \phi}}_{A'B'} \overline{\Sigma}_{ab}^{A'B'}.
\end{equation}
where the symmetric spinor field ${\bf \phi}_{AB}$ is the
Newman-Penrose representation of the self-dual Maxwell field
\cite{pr}.  It is straight forward to re-express the stress-energy
tensor of the Maxwell field,
\begin{equation}\label{ch:stress}
T_{ab} = {1\over 4\pi}\left(
{\bf F}_{ac} {\bf F}_b{}^c - {1\over 4} g_{ab} {\bf F}_{mn}
{\bf F}^{mn} \right)\, ,
\end{equation}
in terms of ${\bf \phi}_{AB}$:
\begin{equation}\label{ch:2}
T_{ab} = - {1\over2\pi} {\bf \phi}_{AB} \overline{{\bf \phi}}_{A'B'} 
\sigma_a^{AA'} \sigma_b^{BB'}.
\end{equation}
The condition $T_{ab} l^a l^b \= 0$ now translates to:
\begin{equation}\label{ch:3}
\o^A \o^B {\bf \phi}_{AB} = 0\, ,
\end{equation}
Or, alternatively, 
\be\label{ch:phi} {\bf \phi}_{AB} = - 2{\bf \phi}_1\i_{(A}\o_{B)} 
+ {\bf \phi}_2 \o_A\o_B 
\ee 
for some (complex) fields ${\bf \phi}_1,{\bf \phi}_2$ on $\Delta$
\cite{pr}.  Note that only the ${\bf \phi}_0 := {\bf \phi}_{AB} \o^A
\o^B$ component ---the ``radiation part''--- of the Maxwell field is
required to be zero at $\Delta$. Since it represent the ``radiation
field'', as one might intuitively expect, the condition $T_{ab} l^a
l^b \=0$ simply states that there is no radiation field at
$\Delta$. It is straightforward to check that the vanishing of other
components of the Ricci tensor does not lead to further restrictions
of the Maxwell field at $\Delta$.

Next, we have to ensure that conditions (v) in the Definition are
satisfied. (v.a) is automatic because the Maxwell stress energy tensor
satisfies the strong energy condition. To ensure that (v.b) is
satisfied, we will require that ${\bf \phi}_1$ is spherically
symmetric on the preferred cross-sections. This restriction is
physically motivated by the fact that, if the radiation field ${\bf
\phi}_0$ were to vanish to the next order at $\Delta$ ---i.e., if $n^a
\nabla_a {\bf \phi}_0 \= 0$--- then the Maxwell field equations {\it
at} $\Delta$ imply that ${\bf \phi}_1$ is spherically symmetric on
$\Delta$.  (Thus, in particular, if there is no electro-magnetic
radiation in a neighborhood  --however small-- of $\Delta$ our
restriction on ${\bf \phi}_1$ will be met.) Now ${\bf \phi}_1$ can be
expressed in terms of the electric and magnetic charges, $Q$ and $P$
of the isolated horizon,
\begin{eqnarray}\label{el-charge}
Q :\= -{1\over 4\pi} \oint_{S_v} {}^*{\bf F} \= {1\over 2\pi} \oint_S\, 
{\rm Re}\,{\bf \phi}_1 \,\, {}^{2}\epsilon \\ 
\label{magn-charge}
P :\= -{1\over 4\pi} \oint_{S_v} {\bf F} \= {1\over 2\pi}
\oint_{S_v} {\rm Im}\, {\bf \phi}_1 \,\, 
{}^{2}\epsilon\, ,
\end{eqnarray}
as follows:
\be
{\bf \phi}_1 \= \frac{2\pi}{a_{\Delta}} (Q + i P)\, .
\ee
(Here $S_v$ are the 2-spheres $v={\rm const}$ in the preferred
foliation. The minus sign in front of the first integrals in
\ref{el-charge} and \ref{magn-charge} arise because we have oriented
$S_v$ such that the radial normal is in-going rather than outgoing.)
In terms of Maxwell fields, this condition can be rewritten as:
\be\label{bcF} 
\pb{{\bF}_{ab}} \= \frac{4\pi iP}{a_{\Delta}}
\,\,\pb{\Sigma_{ab}}^{AB}\, \i_A\, \o_B, \quad {\rm and} \quad
\pb{{}^*{\bF}_{ab}} \= \frac{4\pi iQ}{a_{\Delta}} \,
\pb{\Sigma_{ab}}^{AB}\, \i_A\, \o_B, \ee
Note, however, that there is no restriction on ${\bf \phi}_0$; in
particular, it need not be spherically symmetric. 

Since ${\bf F} = \d{\bf A}$, the magnetic charge $P$ is independent of
the 2-sphere $S_v$ used in its evaluation; indeed we could have used a
2-sphere which does not belong to our preferred family. However, up to
this point the electric charge $Q$ (and hence ${\bf \phi}_1$ and
$\spb{{}^*{\bf F}}$) can be a function of $v$. However, the field
equations satisfied by ${\bf F}$ --condition (iii) in the Definition--
imply that $Q$ is also independent of the choice of the 2-sphere
cross-section: The full content of the field equations is
\be l^a \nabla_a {\bf \phi}_0 \= 0 \quad {\rm and} 
\quad l^a \nabla_a {\bf \phi}_1 \=0  \ee

We are now ready to discuss the action. Recall that, in the gravitational
case, we restricted ourselves to histories in which the area of the
horizon is a fixed constant $a_{\Delta}$. For the same reasons, we will now
restrict ourselves to histories for which the values of electric and
magnetic charges on the horizon are fixed to $Q_{\Delta}$ and $P_{\Delta}$
respectively. To make the action principle well-defined, we need to
impose suitable boundary conditions on the Maxwell fields. Conditions
at infinity are the standard ones. (As in the gravitational case, to
avoid repetition, we will specify them in the phase space framework
which is of more direct interest in this series of papers.)  To find
boundary conditions on $\Delta$, let us consider the standard 
electro-magnetic bulk action:
\begin{equation}
S_{\rm EM} = -{1\over 16\pi} \int_{\cal M} \sqrt{-g} \;
{\bf F}_{ab} {\bf F}^{ab} \d^4x \,\, = 
{1\over 8\pi} \int_{\cal M} {\bf F}\wedge {}^*{\bf F}.
\end{equation}
The numerical factor is adjusted in such a way that the total action
$S_{\rm grav} +S_{\rm EM}$, yields Einstein equations $G_{ab} = 8\pi G
T_{ab}$. Variation of $S_{\rm EM}$ yields
\begin{equation}\label{ch:variation}
\delta\left( {1\over 8\pi} \int_{\cal M} {\bf F}\wedge {}^*{\bf F}
\right) = - {1\over 4\pi} \int_{\cal M} \delta {\bf A} \wedge \delta
{\bf F} + {1\over 4\pi} \int_{\cal \partial M} 
\delta {\bf A}\wedge {}^*{\bf F} \, .
\end{equation}
As usual, the bulk term provides the equations of motion provided the
surface term vanishes. The boundary term (\ref{ch:variation}) vanishes
at infinity due to the fall off conditions.  However, when evaluated
at the horizon, the boundary term (\ref{ch:variation}) does not
automatically vanish. Now, (\ref{ch:phi}) implies that on $\Delta$ the
pull-back of $\bF_{ab}l^b$ vanishes. Therefore, on $\Delta$, the
boundary term in (\ref{ch:variation}) reduces to
\begin{equation}\label{ch:7}
\int \d v\oint_{S_v}\, \delta(\bA \cdot l)\, {}^*F
\end{equation}
where, as before, $v$ is the affine parameter along the integral
curves of $l^a$ such that $v={\rm const}$ define the preferred
foliation of $\Delta$ and $S_v$ are the 2-spheres in this foliation.
Now, since isolated horizons are to be thought of as
``non-dynamical'', it is natural to work in a gauge in which ${\cal
L}_l \spb{{\bf A}_a} \=0$.  Then, (\ref{ch:phi}) implies that ${\bf
A}\cdot l$ is constant on $\Delta$. The form of the boundary term
(\ref{ch:7}) suggests that we fix gauge so that ${\bf A}\cdot l$ is a
{\it fixed} constant on our space of histories. The value of this
constant is then determined by its standard value in the
Reissner-Nordstorm solution:
\be\label{a.l}
{\bf A}_a l^a \= {Q_{\Delta}\over r_{\Delta}}.
\ee
Then the boundary term arising in the variation of the action
vanishes, i.e., the bulk action itself is differentiable and the
action principle is well-defined. Note that the permissible gauge
transformations are now restricted: If $\bA \mapsto \bA + df$, the
generating function $f$ has to (tend to $1$ at infinity and) satisfy
$l^a \partial_a f \= 0$ on $\Delta$. (For further discussion, see
\cite{abf2}).

We will conclude with a summary of the structure of the phase space of
Maxwell fields.  Fix a foliation of ${\cal M}$ by a family of
space-like 3-surfaces $M_t$ (level surfaces of a time function $t$)
which intersect $\Delta$ in the preferred 2-spheres.  Denote by $t^a$
the ``time-evolution'' vector field which is transversal to the
foliation with affine parameter $t$ which tends to a unit
time-translation at infinity and to the vector field $l^a$ on
$\Delta$. In terms of the lapse and shift fields $N$ and $N^a$ defined
by $t^a$, the Legendre transform of the action yields:
%
\be \label{lt2}
S_{\rm EM}=\frac{1}{4\pi}\int \d t \int_{M_t}\,
\left({\bf E}\wedge ({\cal L}_t\bA )-
\d (\bA \cdot t)\wedge {\bf E}-(N\cdot \bF )\wedge{\bf E}-
\frac{N}{2}{}^{(*)}{\bf E}\wedge{\bf E}-\frac{N}{2}{}^{(*)}\bF\wedge
\bF\right)
\ee
where where the 2-form $\bE$ is the pull-back to $M$ of ${}^{*}\bF$,
${}^{(*)}{\bE}_a:=\frac{1}{2}{\epsilon_a}^{bc}{\bE}_{bc}$, and
${}^{(*)}{\bF}_a:=\frac{1}{2}{\epsilon_a}^{bc}{\bF}_{bc}$. {}From
(\ref{lt2}), we can read off the symplectic structure and the
Hamiltonian.

Thus, as usual, the phase space consists of pairs $(\bA, \bE)$ on the
3-manifold $M$, subject to boundary conditions, where the 1-form $\bA$
is now the pull-back to $M$ of the electro-magnetic 4-potential and the
2-form $\bE$ is the dual of the more familiar electric field vector
density. These fields are subject to boundary conditions. On the
horizon 2-sphere $S$, the pull-backs to $S$ of conditions (\ref{bcF})
must hold ensuring that these pull-backs of $\bF$ and $\bE$ are
spherically symmetric. (Since $\bA\cdot l$ appears as a Legendre
multiplier in (\ref{lt2}) condition (\ref{a.l}) does not restrict the
phase space variables at $\Delta$.)  At infinity, $\bA, \bE$ are
subject to the usual fall-off conditions. Consider first the case when
the magnetic charge vanishes so that the vector potential $\bA$ is a
globally defined 1-form on $M$. Then, as usual we require:
\begin{equation}\label{ch:fall-off}
{\bf A} = O\left( {1\over r^{1+\epsilon}}\right), \quad {\rm and} 
\quad
{\bf E} = O\left( {1\over r^2} \right),
\end{equation}
The case when the magnetic charge is non-zero is more subtle since the
vector potential can no longer be specified globally. However, since
we are considering histories with a {\it fixed} magnetic charge $P_{\Delta}$
we can reduce the problem to the one with zero magnetic charge.  Fix,
once and for all, a Dirac monopole potential $\bA_{\Delta}$ with magnetic
charge $P_{\Delta}$ and consider potentials $\bA$ on $M$ of the type
$\bA = \bA_{\Delta} + R$ where the remainder 1-form $R$ has the fall-off $R =
O(1/r^{1+\epsilon}$). For electric fields, we use the same fall-off as
in (\ref{ch:fall-off}). It is straightforward to verify that these
boundary conditions on fields are preserved by evolution equations.
Fields $(\bA, \bE)$ satisfying these conditions constitute our phase
space.

The symplectic structure on this Maxwell phase space can be read off
from (\ref{lt2}):
\begin{equation}
\Omega|_{({\bf A},{\bf E})}\left(\delta_1, \delta_2 \right)
= {1\over 4\pi} \int_M 
\left[ \delta_1 {\bf E}\wedge \delta_2 {\bf A} - 
\delta_2 {\bf E} \wedge \delta_1 {\bf A} \right].
\end{equation}
(The asymptotic conditions ensure that the integrals converge.)
As usual, there is one first class constraint, $\d\bE =0$, which
generates gauge transformations: Under the canonical transformation
generated by $\int f\, \d{\bf E}$, the canonical fields transform, as
usual, via $\bA \mapsto \bA + \d f$ and ${\bf E}$ remains unchanged.
Note that our boundary conditions allow the generating function $f$ to
be non-trivial on the (intersection of $M$ with) $\Delta$; the 
smeared constraint function is still differentiable. Thus, as
in the gravitational case, the gauge degrees of freedom do {\it not} 
become physical in this framework.

{\it Remarks:}

a) It may seem that we could have avoided the fixing of $\bA\cdot l$ on
$\Delta$ and simply introduced instead a boundary term to the action,
involving $\bA\cdot l$. However, then the action would have failed to be
gauge invariant under the local Maxwell gauge transformations.

b) The condition on sphericity of ${\bf \phi}_1$ is somewhat stronger
than the condition on the stress-energy tensor $T_{ab}l^al^b$ (which
is satisfied if and only if $|{\bf \phi}_1|^2$ is spherically
symmetric). However, as pointed out above, the stronger condition is
met if there is no electro-magnetic radiation near $\Delta$.
Furthermore, when ${\bf \phi}_1$ is spherically symmetric, the smeared
Gauss constraint is differentiable even when the generating function
$f$ is non-trivial on the horizon $S$ so that local gauge
transformations on the horizon are regarded as `gauge' also in the
Hamiltonian framework.  If we had imposed only the weaker condition,
it would have been awkward to sort out the degrees of freedom in the
Hamiltonian framework.

\subsection{Dilatonic Couplings}
\label{sec5.2}

In this subsection we add a dilatonic charge to the Einstein-Maxwell
theory, i.e., consider a scalar field which is coupled non-minimally
to the Maxwell field. (Further details can be found in \cite{30}.)

The dilatonic field theory has a bulk contribution to the action of
the form,
\be
S_{\rm Dil}=-\frac{1}{16\pi}\int_{{\cal M}}\sqrt{-g} [2(\nabla\phi)^2
+ e^{-2\alpha\phi}\bF_{ab}\bF^{ab}]\d^4x\label{dil:act}
\ee
where $\alpha$ is a free parameter which governs the strength of the
coupling of the dilaton to the Maxwell field. When $\alpha=0$ we
recover the Einstein-Maxwell-Klein-Gordon system, while for $\alpha=1$
$S_{\rm Dil}$ is part of the low energy action of string theory.  The
total action is the dilatonic action plus the gravitational part
considered in Section \ref{sec3}.

For the convenience of readers who may not be familiar with dilatonic
gravity, we will first recall a few releyvant facts.  The standard
equations of motion that follow from $S_{\rm Dil}$ are:
\ba
\nabla_a(e^{-2\alpha\phi}\bF^{ab})=0,\label{dil:eom1}\\
\nabla^2\phi+\frac{\alpha}{2}e^{-2\alpha\phi}\bF^2=0,
\label{dil:eom2}\\
R_{ab}=2\nabla_a\phi\nabla_b\phi+2e^{-2\alpha\phi}\bF_{ac}{\bF_b}^c
-\frac{1}{2}
g_{ab}e^{-2\alpha\phi}\bF^2,\label{dil:eom3}
\ea
where $\bF^2=\bF_{ab}\bF^{ab}$. The first equation (\ref{dil:eom1})
can be rewritten as $\d (e^{-2\alpha\phi}\,{}^*\bF)=0$, so there is a
conserved charge,
\be
\tilde{Q}:=\frac{1}{4\pi}\oint_{S}e^{-2\alpha\phi}{}^*\bF \ee
where $S$ is any 2-sphere `containing the black hole' (the integral is
independent of the sphere as long as they are homologous). we will
refer to it as the {\it dilatonic charge} to distinguish it from the
electric charge
\be \label{Q}
Q_S := \frac{1}{4\pi} \oint_S {}^*F . \ee
We will see that our boundary conditions directly imply that the
electric charge is conserved along the horizon. Thus, by choosing $S$
to be a 2-sphere cross-section of the horizon, we obtain a charge,
$Q_{\rm bh}$, which is intrinsically associated with the black hole.
Next, note that Eq.(\ref{dil:eom1}) can be rewritten as $\d
{}^*\bF=2\alpha\d \phi\wedge{}^*\bF$. Therefore, we obtain a current
three-form and a conserved quantity: Since
\be
\int_{M}(\d {}^*\bF-2\alpha\d \phi\wedge{}^*\bF) = 0
\label{dil:curr}
\ee
for any  (partial) Cauchy surface $M$, the difference
\be
Q_{\infty}-Q_{\rm bh}=\frac{\alpha}{2\pi}
\int_{M}\d \phi\wedge{}^*\bF
\ee
between the electric charge $Q_{\infty}$ measured at spatial infinity
and $Q_{\rm bh}$ evaluated at the intersection of $M$ and $\Delta$ is
also conserved.

Let us now impose boundary conditions to ensure that we have a
well-defined action principle for dilatonic and Maxwell fields. At
infinity we will impose the same boundary conditions for the Maxwell
field as in Section \ref{sec5.1}. For the dilaton field we will require that
it tend to a constant, $\phi_{\infty}$, at infinity, and that its
derivatives fall-off as $O(1/r^2)$. These conditions suffice for
vanishing of the boundary terms at infinity that result from the
variation of action. 

Let us now turn to the boundary conditions at the horizon, starting
with (v.a).  We can read-off the stress-energy tensor from Eq
(\ref{dil:eom3}) and verify that $T_{ab}l^al^b \ge 0$. The
Raychaudhuri equation now implies that $R_{ab}l^al^b \=0$.  Therefore,
transvecting (\ref{dil:eom3}) with $l^al^b$, we conclude that
$\dot{\phi}:=l^a\nabla_a\phi \=0$, {\it and} that the electro-magnetic
field tensor is of the form (\ref{ch:phi}). Next, it is simple to
verify that the condition that $T_{ab}l^b$ be causal implies that
$m^a\nabla_a \=0$. There the dilaton field $\phi$ is constant on the
horizon. It is easy to check that the constancy of $\phi$ is also
sufficient to ensure that $T_{ab}l^b$ be causal. Next, consider
condition (v.b). Since $\phi$ is constant on $\Delta$, as in the
Maxwell case, (v.b) can be satisfied by demanding that the component
$\phi_1$ of the Maxwell field be spherically symmetric. Finally, note
that since the dilaton is $v$ independent on the horizon equations
(\ref{ch:phi}),(\ref{dil:curr}) imply that $\int_{\Delta}\d{}^*\bF=0$,
so the electric charge $Q_{\rm bh}$ is independent of the 2-sphere of
integration in (\ref{Q}).  As before, we will restrict our histories
to have specific values of area and of charges $\tilde{Q}$ and $Q_{\rm
bh}$: $a_\Delta, \tilde{Q}_\Delta$ and $Q_\Delta$. (For simplicity, in
this section we have set the magnetic charge $P$ to zero.)

We are now ready to analyze the differentiability of $S_{\rm Dil}$.
The variation of the action (\ref{dil:act}) yields two surface terms,
one from the variation of the Maxwell field and the other from the
variation of the dilaton field. Both vanish at infinity because of the
standard boundary conditions. At the horizon, the first term gives:
\be
\frac{e^{-2\alpha\phi}}{4\pi}
\int_{\Delta}\,\delta \bA\wedge{}^*{\bf F}
\ee
As in section (\ref{sec5.1}), we will gauge fix $\bA\cdot l$ to its
standard value in the static solution. Then $\delta\bA\cdot l =0$
and the integrand in the boundary term vanishes because the Maxwell
field has the form (\ref{bcF}). 

Let us now consider the variation of $S_{\rm Dil}$ with respect to the
dilaton field. The corresponding boundary term is
\be
\frac{1}{4\pi}\int_{{\Delta}}{}^*\d\phi\;\delta\phi
\ee 
This term also vanishes since $\dot{\phi}$ vanishes on $\Delta$.
Thus, with our boundary conditions, the bulk action $S_{\rm Dil}$ is
differentiable by itself.

Next, let us examine the structure of the phase space of the
theory. Consider first the the term $S_{\rm EM}=-\frac{1}{16\pi}\int
\d^4x \sqrt{-g}e^{-2\phi}\bF^2$ in the action $S_{\rm
Dil}$. Performing the usual $3+1$ decomposition and defining a time
translation vector field $t^a$ that tends to $l^a$ at the horizon, we
find that the canonically conjugated momenta $\bf {\Pi}^a$ is 
given by:
\be
{\bf{\Pi}}^a=\frac{\sqrt{h}}{4\pi}e^{-2\alpha\phi}{\bf E}^a
\ee
where $h$ is the determinant of the metric $h_{ab}$ on $M$ and ${\bf
E}^a$ is the usual electric field ${\bf E}^a=g^{ab}\bF_{cb}{\bf n}^c$.
The action then reads,
\ba
&{}& S_{\rm EM} = \frac{1}{4\pi} \int \d t \int_{M_t} \,
e^{-2\alpha \phi} \\ \nonumber
&{}& \left[
{\bf E}\wedge ({\cal L}_t\bA) -
\d (\bA \cdot t)\wedge {\bf E}-(N\cdot \bF )\wedge{\bf E}-
\frac{N}{2}{}^{(*)}{\bf E}\wedge{\bf E}-\frac{N}{2}{}^{(*)}
\bF\wedge \bF\right],
\ea
where, as before, ${}^{(*)}$ denotes the 3-dimensional dual so that
${}^{(*)}{\bf E}_a:=\frac{1}{2}{\epsilon_a}^{bc}{\bf E}_{bc}$ and
${}^{(*)}{\bf F}_a:=\frac{1}{2}{\epsilon_a}^{bc}{\bf F}_{bc}$.

The $3+1$ decomposition of the term $S_{\phi}=-\frac{1}{8\pi}\int
\d^4x \sqrt{-g}(\nabla\phi)^2$, gives as the momentum conjugate to
$\phi$,
\be
{\Pi}=\frac{1}{4\pi}\sqrt{h}\; {\bf n}^a\nabla_a\phi.
\ee
where ${\bf n}^a$ is the unit normal to $M$, and the action can be written as
\be
S_{\phi}=\int \d t\int_{M}\d^3x\left[{\Pi}\dot{\phi}-
\frac{2\pi N}{\sqrt{h}}
{\Pi}^2-N^a(\nabla_a\phi) {\Pi}+\frac{\sqrt{h}}{2\pi N}(N^a
\nabla_a\phi)^2+\frac{1}{8\pi}h^{ab}\nabla_a\phi\nabla_b\phi\right]
\ee

Thus, the matter phase space consists of pairs $({\bf A},{\bf \Pi})$
and $(\phi,\Pi)$ on $M$, satisfying our boundary conditions.  The
symplectic structure can be written as
\begin{equation}
\Omega|_{({\bf A},{\bf \Pi})}\left(\delta_1, \delta_2\right)
= {1\over 4\pi} \int_M 
\left[ \delta_1 {\bf \Pi}^a \delta_2 {\bf A}_{a} - 
{\delta_2 {\bf \Pi}^a} \delta_1 {\bf A}_a \right].
\end{equation}
for the Maxwell part and,
\begin{equation}
\Omega|_{(\phi,\Pi)}\left(\delta_1, \delta_2\right) 
= {1\over 4\pi} \int_M 
\left[ \delta_1 {\Pi} \delta_2 {\phi} - 
{\delta_2 {\Pi}} \delta_1 {\phi} \right].
\end{equation}
for the dilaton.  The gauge transformations in the Hamiltonian
framework are the Maxwell $\U(1)$ gauge rotations. The Hamiltonian can
be read off from the Legendre transforms of the action given
above. 

%
%

\section{Discussion}
\label{sec6}

In this paper, we introduced the notion of a non-rotating isolated
horizon, constructed an action which yields Einstein's equations for
histories admitting these horizons, and obtained a Hamiltonian
framework in terms of real $\su(2)$-valued fields. We found that there
is a surprising and interesting interplay between the space-time
boundary conditions satisfied by isolated horizons $\Delta$ and the
Chern-Simons theory for a ${\U}(1)$ connection on $\Delta$. The
gravitational part of boundary conditions, the corresponding action
and the phase space framework could be discussed in the general case,
without committing oneself to specific matter fields. In addition, we
discussed in detail the boundary conditions, action and the
Hamiltonian formulation for Maxwell and dilaton fields. This detailed
framework has five parameters:%
\footnote{In the standard treatments, one uses the ADM mass $M$ as the
fundamental parameter and expresses area $a_\Delta$ in terms of $M$
and other charges. By contrast, we regard $a_\Delta$ as the basic
parameter and $M$ as a secondary quantity derived from Hamiltonian
considerations \cite{abf1,abf2}. Thus, all our parameters --including
charges in the dilatonic case-- are defined directly at the horizon,
without reference to infinity. This is particularly important for
cosmological horizons where $M$ may not even be defined! In this case,
we will be still be able to account for the entropy \cite{5}.}%
the cosmological constant $\Lambda$, the horizon area $a_\Delta$, the
electric and magnetic charges $Q$ and $P$, and the dilaton charge
$\tilde{Q}$.  Since our primary motivation is to present a Hamiltonian
framework that will serve as a point of departure for the entropy
calculation of a quantum black hole in \cite{5}, in this paper we
worked with sectors of the theory in which all these charges are
fixed. A more general discussion is necessary to extend the laws of
black hole mechanics to isolated horizons and is given in \cite{abf2}.

In the final phase space framework, the basic variables consist of
pairs, $(\Ag, \Eg)$ of $\su(2)$-valued forms as in the simpler case
without internal boundaries. However, they are now subject to boundary
conditions not only at infinity but also on the internal boundary $S$
representing the isolated horizon: on $S$, the independent information
in $\sdpb{\Ag}$ is contained in the ($\gamma$ independent) ${\rm
U}(1)$ connection $V$ whose curvature ${\cal F}$ is proportional to
$\sdpb{\Eg}^{AB}\i_A\o_B$ (see (\ref{calF})).  Of particular interest
is the interplay between various fields at $S$. In the gravitational
sector, the boundary conditions fix neither the soldering form
$\sigma$ nor the connection $A$ on $S$ but rather a {\it relation}
between them, namely (\ref{calF}). As a consequence, quantum theory
will allow fluctuations in both these fields but they will be
intertwined by the quantum analog of (\ref{calF}). In the matter
sector, on the other hand, the dynamical fields on $S$ are completely
determined by the values of charges and geometry ---more precisely,
the field $\sdpb{\Sigma}$--- of $S$. This will turn out to be the key
reason why the entropy depends only on the area $a_\Delta$ of $S$,
independently of the matter content.  Finally, we saw that the
presence of the horizon modifies symplectic structure: in addition to
the standard bulk term, there is now a surface term which coincides
with the symplectic structure of a ${\U}(1)$ Chern-Simons theory for
$V$ (see (\ref{Omega})). This feature will also play a key role in the
quantization of the theory in \cite{5}.

The isolated horizons introduced here are special cases of Hayward's
trapping horizons \cite{9} in that we require that the expansion of
$l$ should vanish (and $n$ to satisfy certain restrictions.) This
condition on $l$ was necessary for us because, for entropy
considerations, we wish to focus on isolated horizons. While there are
several similarities between the two sets of analyses, there are also
important differences in the motivation and hence also in the
subsequent developments. Hayward's papers discuss dynamical situations
and, since he is specifically interested in characterizing black
holes quasi-locally, he makes a special effort to rule out
cosmological horizons. By contrast, we are primarily interested in the
interface of general relativity, statistical mechanics and quantum
field theory. Therefore, we focus on equilibrium situations and wish
to incorporate not only black holes but also cosmological horizons (as
well as other situations such as those depicted in figure 2) which
are of importance in this context. Finally, because of our further
restrictions on $l$ and $n$, we were able to obtain an action
principle and construct a detailed Hamiltonian framework which is
necessary for quantization.

There are two directions in which our framework could be extended.
First, one should weaken the boundary conditions and carry out the
subsequent analysis to allow for rotation. As indicated in section
\ref{sec3.1}, only conditions (iv.b) and (v.b) in the main Definition
have to be weakened. Recent results of Lewandowski \cite{jl}
have paved way for this task. Second, already in the non-rotating
case, one could allow horizons which are distorted, e.g., because the
presence of a cage around the black hole. It may not be possible to
introduce an action principle and a complete Hamiltonian framework in
this case since the matter fields involved are not fundamental.
However, it should still be possible to introduce appropriate boundary
conditions, work out their consequences and explore generalizations of
the laws of black hole mechanics. The second type of extension
involves going from general relativity to more general, higher
derivative theories. In these cases, the boundary conditions at
$\Delta$ would be the same but there would be some differences in
their consequences arising from the differences in the field
equations. The action principle and the Hamiltonian framework would be
significantly different. However, it should be possible to carry out
the required extension in a tetrad framework \cite{abf2}. This would
provide an interesting generalization of Wald's analysis \cite{wald}
based on Noether charges of stationary solutions.

\section{Acknowledgments} We are most grateful to John Baez, 
Chris Beetle and Steve Fairhurst for many stimulating discussions and
to Jerzy Lewandowski for communicating his results prior to
publication.  We have also profited from comments made by numerous
colleagues, especially Brandon Carter, Sean Hayward, Don Marolf,
Daniel Sudarsky, Carlo Rovelli, Thomas Thiemann and Robert Wald.  The
authors were supported in part by the NSF grants PHY94-07194,
PHY95-14240, INT97-22514 and by the Eberly research funds of Penn
State.  In addition, KK was supported by the Braddock fellowship of
Penn State. AC was also supported by DGAPA-UNAM grants IN106097,
IN121298 and by CONACyT Proy. Ref. I25655-E.

\appendix
\section{Conventions}
\label{appA}

In this paper primed and unprimed upper case letters stand for
$\SL(2,C)$ spinor indices, lower case letters denote space-time tensor
indices. (In sections on the phase space framework, the unprimed upper
case letters denote the $\SU(2)$ spinor indices and the lower case letters
denote spatial indices.) The conventions are the same as in
\cite{18}. These differ somewhat from the Penrose-Rindler
\cite{pr} conventions because while their metric has signature
+,-,-,-  ours has signature -,+,+,+.

The soldering form (or, equivalently, the tetrad field) defines the
metric via
\begin{equation} \label{metric}
g_{ab} = \sigma_a^{AA'}\sigma_{b AA'}.
\end{equation}
The soldering form $\sigma$ is required to be anti-hermitian
($\overline{\sigma}_a^{AA'} = -\sigma_a^{A'A}$) so that $g$ defined
through (\ref{metric}) is a real Lorentzian metric of signature
$(-+++)$. The self-dual connection $A$ defines a derivative operator
$D_a$ that operates on {\it unprimed} spinors and defines the
connection 1-form $A$ via $D_a\lambda_A=\partial_a\lambda_A + A_{a
A}^{\,\,\,\,\,B}\lambda_B$.

The self-dual 2-forms $\Sigma$ are defined by:
\begin{equation}\label{sigma}
\Sigma_{ab}^{AB} = 2 \sigma_{[a}^{AA'} \sigma_{b]\,\,A'}^{B} = 
2 \sigma_a^{(A|A'|} \sigma_{b\,\,A'}^{B)}.
\end{equation}
The spinorial equivalents of the null tetrad are given by 
\begin{eqnarray}
l^a &=& i\,\o^A \o^{A'} \sigma^a_{AA'} \label{l} \\
n^a &=& i\,\i^A \i^{A'} \sigma^a_{AA'} \label{n} \\
m^a &=& \o^A \i^{A'} \sigma^a_{AA'}    \label{m} \\
\overline{m}^a &=& - \i^A \o^{A'} \sigma^a_{AA'} \label{mbar}
\end{eqnarray}

Our choice of orientation is as follows. The volume 4-form 
in ${\cal M}$ is taken to be
\begin{equation}\label{orient}
\epsilon_{abcd} = 24i l_{[a} n_b m_c \overline{m}_{d]};
\end{equation}
the volume 3-form on $M$ is 
\be
\epsilon_{abc} = \epsilon_{abcd}{\tau}^d\, ;
\ee
while that on $\Delta$ is 
\be 
{}^\Delta\!\epsilon_{abc} = -6i n_{[a} m_b \overline{m}_{c]}\, .
\ee
Finally, the area 2-form on $S$ is assumed to be
\begin{equation}\label{2-orient}
\epsilon_{ab} = 2i m_{[a} \overline{m}_{b]},
\end{equation}

The Newman-Penrose components of the Weyl tensor are given by:
\begin{eqnarray}
\Psi_0 &=& \Psi_{ABCD} \o^A\o^B\o^C\o^D = - C_{abcd} l^a m^b l^c m^d
\nonumber \\ 
\Psi_1 &=& \Psi_{ABCD} \o^A\o^B\o^C\i^D = i C_{abcd} l^a
m^b l^c n^d \nonumber \\ 
\Psi_2 &=& \Psi_{ABCD} \o^A\o^B\i^C\i^D =
C_{abcd} l^a m^b \bar{m}^c n^d \label{weilscalars} \\ 
\Psi_3 &=& \Psi_{ABCD} \o^A\i^B\i^C\i^D = -i C_{abcd} l^a n^b 
\bar{m}^c n^d \nonumber \\ 
\Psi_4 &=& \Psi_{ABCD} \i^A\i^B\i^C\i^D = - C_{abcd}
\bar{m}^a n^b \bar{m}^c n^d, \nonumber
\end{eqnarray}
where $C_{abcd}$ is the Weyl tensor. Finally, the tetrad components of
the Ricci tensor are given by:

\ba
\Phi_{00} &= \Phi_{ABA'B'} \o^A\o^B\o^{A'}\o^{B'} &= \frac{1}{2}
R_{ab} l^a l^b\nonumber\\ 
\Phi_{01} &= \Phi_{ABA'B'} \o^A\o^B\o^{A'}\i^{B'} &= i \frac{1}{2}
R_{ab} l^a m^b \nonumber\\
\Phi_{02} &= \Phi_{ABA'B'} \o^A\o^B\i^{A'}\i^{B'} &= -\frac{1}{2}
R_{ab} m^a m^b; \nonumber\\
\Phi_{10} &= \Phi_{ABA'B'} \o^A\i^B\o^{A'}\o^{B'} &= -i \frac{1}{2}
R_{ab} l^a \bar{m}^b; \nonumber\\
\Phi_{11} &= \Phi_{ABA'B'} \o^A\i^B\o^{A'}\i^{B'} &= 
\frac{1}{4}R_{ab} (l^a n^b+m^a \bar{m}^b) ; \label{ricciscalars} \\
\Phi_{12} &= \Phi_{ABA'B'} \o^A\i^B\i^{A'}\i^{B'} &= 
i \frac{1}{2}R_{ab} n^a m^b; \nonumber\\ 
\Phi_{20} &= \Phi_{ABA'B'} \i^A\i^B\o^{A'}\o^{B'} =& 
-\frac{1}{2}R_{ab} \bar{m}^a \bar{m}^b; \nonumber\\
\Phi_{21} &= \Phi_{ABA'B'} \i^A\i^B\o^{A'}\i^{B'} =& 
-i \frac{1}{2}R_{ab} n^a \bar{m}^b; \nonumber\\
\Phi_{22} &= \Phi_{ABA'B'} \i^A\i^B\i^{A'}\i^{B'} =& 
\frac{1}{2}R_{ab} n^a n^b; \nonumber\\
\ea
(As usual, $\Phi_{ab} := \sigma_a^{AA'}\, \sigma_b^{BB'}\Phi_{AA'BB'}$
is the traceless part of the Ricci tensor: $-2\Phi_{ab}= R_{ab} -
\textstyle{R\over 4} g_{ab}$.)

Finally, the Newman-Penrose components of the Maxwell field are
given by:
\ba 
{\bf \phi}_0 &= {\bf \phi}_{AB} \o^A \o^B &= -i F_{ab}l^am^b\nonumber\\
{\bf \phi}_1 &= {\bf \phi}_{AB}i^A \o^B &= -{1\over 2} F_{ab}(l^an^b
- m^a\overline{m}^b) \nonumber\\
{\bf \phi}_2 &= {\bf \phi}_{ab}\i^A \i^B &= -i F_{ab}n^a\overline{m}^b
\nonumber\\
\ea
\section{Some consequences of boundary conditions}
\label{appB}

In this appendix we sketch proofs of assertions made in section
\ref{sec3.2}. An alternate procedure, tailored to the Newman Penrose
framework, and other consequences of the boundary conditions
which are not directly needed here can be found in \cite{abf2}.

1. To see that the Lie derivative of the induced metric on $\Delta$
with respect to $l$ is zero, note first that the condition (iv.a)
implies%
\footnote{As in the main text, note that the pull-back operation 
is redundant for covariant fields which are defined intrinsically 
on $\Delta$.}
\begin{equation}\label{d\o}
\spb{D_a} \o_A \= - {\alpha_a} \o_A
\end{equation}
for some one-form $\alpha_a$ on $\Delta$. Now, using the expression
(\ref{l}) of $l$ in terms of the spinors and the compatibility of
$\sigma$ and $A$ implied by condition (iii) of the main Definition,
one can easily find the expression for $\spb{\nabla_a}\, l_b$ in terms of
$\alpha_a$:
\be\label{app:dl}
\spb{\nabla_a} l_b \= - 2{U_a} l_b\, ,
\ee
where, as in the main text, $U_a \= {\rm Re} \alpha_a$. (Conditions
(iv.b) on $n$ imply that $U_a = -(U\cdot l)n_a$.) Eq (\ref{app:dl})
implies that $l$ is a geodesic vector field,
\[
l^a \nabla_a l_b \= -2 (l^a U_a) l_b,
\]
and that Lie derivative with respect to $l$ of the metric induced on
$\Delta$ vanishes:
\be
{\cal L} \pb{g_{ab}} \= 2 \nabla\pb{{}_{(a} l_{b)}} = 0.
\ee
It also implies that $l$ is twist-free, shear-free and
expansion-free. In fact, the condition (iv.a) is equivalent
to the requirement that $l$ be geodesic, twist-free, torsion-free and
expansion-free.

2. Next, let us show that the connection $A$ on $\Delta$ is of the
form (\ref{A}). Eq (\ref{d\o}) tells us a constraint on the connection
imposed by the condition (iv.a) of the main Definition. Condition
(iv.b) of the Definition provides a further restriction:
\begin{equation} \label{d\i}
\spb{D_a}\, \i_A \= {\alpha_a}\, \i_A + i f(v) {\bar{m}_a}\, 
\o_A,
\end{equation}
where $\alpha_a$ is the same 1-form as in (\ref{d\o}) (because the
spin dyad is normalized $\i^A \o_A \= 1$), and $f(v)$ is a function of
$v$ only as introduced by the boundary condition (iv.b). A general
connection $A_a^{AB}$ acting only on unprimed spinors can be
decomposed into its $\i^{(A}\o^{B)}, \i^A\i^B, \o^A\o^B$
components. Then, in the gauge adapted to the dyad, $\partial_a \i^A
\= \partial_a \o^A \= 0$, the connection has only $\i^{(A}\o^{B)},
\o^A\o^B$ components, as in (\ref{A}): \be \spb{A_a}^{AB} \=
-2\alpha_a\i^{(A}\o^{B)} - \beta_a \o^{A}\o^{B} \ee
It now follows immediately that the pull-back to $\Delta$ of the curvature 
is given by:
\be\label{F1}
\pb{F_{ab}}^{AB} = -4 (\partial_{[a}\alpha_{b]})\, \i^{(A}\o^{B)}
- 2(\partial_{[a}\beta_{b]} - 2\alpha_{[a}\beta_{b]})\, \o^A\o^B 
\ee
We will provide a proof of the assertions (\ref{mu}) and (\ref{beta})
on properties of $U$ and $\beta$ after discussing the properties of
curvature tensors (see 5. below).

Finally, let us establish (\ref{nu=gamma}). We assume that the slice
$M$ intersects $\Delta$ in $S$ such that on $S$ the unit normal
$\tau^a$ to $M$ is given by $\tau^a = (l^a+n^a)/\sqrt{2}$. Therefore,
using (\ref{d\o}) and (\ref{d\i}), we can compute the part
$\sdpb{K_a}{}^{AB}$ of the extrinsic curvature of $M$ at points of $S$.
One obtains:
\be \sdpb{K_a}{}^{AB} = {f(v)\over\sqrt{2}}(m_a\i^A\i^B +
\overline{m}^a \o^A\o^B)
\ee
Since the connection $A$ on $M$ is given by: $A_a{}^{AB} =
\Gamma_a{}^{AB} -{i\over \sqrt{2}}K_a{}^{AB}$, using the definition of
$V$ we have the required result:
\be \sdpb{V_a} = -i \sdpb{\Gamma_a}{}^{AB}\i_A\o_B\, . \ee

3. As is well-known, the full Riemann curvature can be decomposed into
its (self- and antiself-dual) Weyl tensor, traceless Ricci tensor, and
scalar curvature. In the spinorial notation, we have (see, e.g.,
\cite{pr}):
\begin{eqnarray}\label{R}
R_{AA'BB'CC'DD'} = \Psi_{ABCD} \epsilon_{A'B'} \epsilon_{C'D'}
+ \bar{\Psi}_{A'B'C'D'} \epsilon_{AB} \epsilon_{CD} \nonumber \\
+ \Phi_{ABC'D'} \epsilon_{A'B'} \epsilon_{CD}
+ \bar{\Phi}_{A'B'CD} \epsilon_{AB} \epsilon_{C'D'} \label{decomposition} \\
+ {R\over 12}(\epsilon_{AC}\epsilon_{BD}\epsilon_{A'C'}\epsilon_{B'D'}
- \epsilon_{AD}\epsilon_{BC}\epsilon_{A'D'}\epsilon_{B'C'}). \nonumber
\end{eqnarray}
The Weyl spinor $\Psi_{ABCD}$ and the trace-free Ricci spinor
$\Phi_{AA'BB'}$ can be further expanded in terms of their components
in the spinor basis $\i^A, \o^A$ defined in appendix \ref{appA},

Let us consider the Ricci tensor. Since $T_{ab}l^a l^b\ge 0$ the
Raychaudhuri equation for $l^a$ implies $T_{ab}l^al^b \=0$, whence,
via Einstein's equation which holds on $\Delta$, we conclude
$\Phi_{00} \= 0$. Since $T_{ab}l^b$ is causal and $T_{ab} l^a l^b
\=0$, it follows that $T_{ab}m^al^b \=0$, whence we have: $\Phi_{10}
\=0$ and $\Phi_{01} \=0$. These conditions can be summarized in the
following equation:
\begin{eqnarray} \label{ricci}
\Phi_{ABA'B'} \= &4 \Phi_{11} \i_{(A}\o_{B)} \i_{(A'}\o_{B')}
+\Phi_{22} \o_A\o_B\o_{A'}\o_{B'}
+\Phi_{02} \i_A\i_B\o_{A'}\o_{B'} \\ \nonumber
& +\Phi_{20}\o_{A}\o_{B}\i_{A'}\i_{B'} - 2\Phi_{12}\i_{(A}\o_{B)}
-2 \Phi_{21} \o_A\o_B \i_{(A'}\o_{B')}.
\end{eqnarray}
(We will show later that $\Phi_{20}$ and $\Phi_{02}$ also vanish.)

Let us now turn to the Weyl tensor. Using the equation (\ref{app:dl})
one gets: 
\be\label{app:ddl} 
\pb{R_{ab}}{}_{cd}l^d =  2\spb{\nabla_{[a}}\spb{\nabla_{b]}} l_c
= -2(\nabla_{[a}\mu_{b]}) l_c \=0, \ee 
Transvecting this equation with appropriate vectors and using the fact
that the trace of the Weyl tensor vanishes, we conclude that
\be \Psi_0 \= 0\quad {\rm and} \quad \Psi_1 \=0. \ee
Next, using the boundary conditions (iv.a), (iv.b), one
can calculate the derivative of $n_a$:
\be\label{app:dn}
\spb{\nabla_a} n_b = 2U_a n_b - 2 f m_{(a}\bar{m}_{b)}.
\ee
Hence, it follows that
\ba \label{app:ddn}
\pb{R_{ab}}{}_{cd}n^d = 2\spb{\nabla_{[a}} \spb{\nabla_{b]}}n_c = -
4f(\bar{m}_{[a} U_{b]}m_c + m_{[a} U_{b]}\bar{m}_c) - 
(\partial_{[a} f) m_{b]} \bar{m}_c - 
(\partial_{[a} f) \bar{m}_{b]} m_c.
\ea
Transvecting this equation with suitable vectors and using the
trace-free property of the Weyl tensor and the fact that we have set
$f = 1/r_{\Delta}$, we conclude:
\be \label{psi2,psi3}
{\rm Im}\Psi_2 \= 0, \quad \Psi_2 + \frac{R}{12} \= 
{2U\cdot l\over r_\Delta},
\quad {\rm and} \quad \Psi_3 - \Phi_{21} \= 0 
\ee
Consequently, the Weyl spinor has the form:
\be \label{weyl}
\Psi_{ABCD} \= 6 \Psi_2 \i_{(A}\i_B\o_C\o_{D)} 
- 4 \Psi_3 \i_{(A}\o_B\o_C\o_{D)}
+ \Psi_4 \o_A\o_B\o_C\o_D.\ee
where $\Psi_2$ and $\Psi_3$ are subject to (\ref{psi2,psi3}).

Next, Recall \cite{18} the expression for $F$ in terms of the
self-dual part of the Riemann curvature:
\be
F_{ab}^{AB} = -{1\over4} R_{ab}^{\quad cd} \Sigma_{ab}^{AB}.
\ee
Using the decomposition (\ref{R}) of the Riemann tensor, we obtain:
%
\be\label{F2}
F_{ab\,CD} \= -{1\over2}\Psi_{ABCD}\Sigma_{ab}^{AB}
-{1\over2}\bar{\Phi}_{A'B'CD}\bar{\Sigma}_{ab}^{A'B'}
-{R\over 24}\Sigma_{ab\,CD}.
\ee
Finally, using the identity:
\begin{equation}\label{app2:pbs}
\pb{\Sigma_{ab}}^{AB} \= 4\i^{(A}\o^{B)} m\pb{{}_{[a}\bar{m}_{b]}}
+ 4i \o^A\o^B n\pb{{}_{[a}\bar{m}_{b]}},
\end{equation}
which follows from the definitions of the null tetrad, and equations
(\ref{ricci}) and (\ref{weyl}), we can express the pull-back
$\pb{F_{ab}}^{CD}$ of $F$ to $\Delta$ as:
\be \label{pbF}
\pb{F_{ab}}^{CD} \= (\Psi_2-\Phi_{11}-{R\over 24})
\pb{\Sigma_{ab}}^{CD} 
- 2i (3\Psi_2-2\Phi_{11}) \o^C\o^D n\pb{{}_{[a}\bar{m}_{b]}}.
\ee
where we have used $\Psi_3 - \Phi_{21} =0$. (Other Weyl and Ricci
components do not appear because $F$ is pulled-back to $\Delta$.)
This equation can also be rewritten in the following simpler
form:
\be
\pb{F_{ab}}^{AB} \= \left [
(\Psi_2-\Phi_{11}-{R\over 24})\delta^A_C \delta^B_D
-({3\Psi_2\over2}-\Phi_{11})\o^A\o^B\i_C\i_D \right] 
\pb{\Sigma_{ab}}^{CD}.
\ee

Next, we will use the Bianchi identity $\D\wedge F \= 0$ and the
equation of motion $\D\wedge\Sigma \=0$ to extract further information
about $\Psi_2, \Phi_{11}$ and $R$ which appear in the above relation
between $\spb{F}$ and $\spb{\Sigma}$. Using (\ref{d\o}) and
(\ref{d\i}), we obtain:
\ba \label{app:df}
0 \= \spb{\nabla_{[a}} \pb{F_{bc]}}^{AB} = \big[\,
\pb{\nabla_{[a}}\,{}(\Psi_2-\Phi_{11}- {R\over 24})\delta^A_C 
\delta^B_D
-\pb{\nabla_{[a}}{}({3\Psi_2\over2}-\Phi_{11})\,\o^A\o^B\i^C\i^D
\nonumber\\ 
- 2({3\Psi_2\over2}-\Phi_{11})\,\spb{\beta_{[a}}\o^A\o^B\o_C\i_D\, 
] \pb{\Sigma_{bc]}}^{CD}. 
\ea
Transvecting this equation with $\o_A\o_B$, we get an identity.
Transvecting it with $\i_A\o_B$ we conclude: $l^a\partial_a (\Psi_2
-\Phi_{11} -\frac{R}{24}) \= 0$. Transvecting it with $\i_A\i_B$, we
conclude that $\Psi_2 + \frac{R}{12}$ is spherically symmetric.
Recall furthermore that by condition (v.b) in the main Definition,
$T_{ab}l^an^b$ is spherically symmetric. Using Einstein's equation on
$\Delta$ we conclude that $(\Psi_2 -\Phi_{11} - \frac{R}{24})$
is spherically symmetric on $\Delta$. 

Finally, the pull-back of (\ref{pbF}) to any 2-sphere in our preferred
foliation yields:
\be \label{F3}
\dpb{F_{ab}}^{CD} \= (\Psi_2-\Phi_{11}-\frac{R}{24})\, 
\dpb{\Sigma_{ab}}^{CD}
\ee
We can transvect this equation with $\i_C\o_D$ and integrate the
result on a 2-sphere. The left side is then $-2\pi i$, the Chern number
of the spin-bundle of the $U(1)$ connection, while the factor in the
parenthesis on the right side, being constant, comes outside the
integral and the integral itself then yields just
$ia_{\Delta}$. Hence, we have:
\be\label{app:const}
\Psi_2-\Phi_{11}-{R\over 24} \= - {2\pi\over a_{\Delta}}.
\ee
An alternate (non-spinorial) way to obtain this result is to note that
the left side of (\ref{app:const}) equals $- ({}^2R/4)$ where ${}^2R$ is
the scalar curvature of the 2-sphere cross-sections \cite{pr}. Then,
using the Gauss Bonnet theorem, we obtain (\ref{app:const}).

5. It only remains to show the properties (\ref{mu}) and (\ref{beta})
of $\mu$ and $\beta$ and vanishing of $\Phi_{20}$ and $\Phi_{02}$.

Eq (\ref{mu}) is an immediate consequence of the
second equation in (\ref{psi2,psi3}) and Einstein's equation on
$\Delta$. To obtain the first equation on $\beta$, let us transvect
(\ref{F1}) with $\i_A\i_B$. Using (\ref{pbF}), one can express the left
side as
\be \label{F4}
\pb{F_{ab}}^{AB}\i_A\i_B \=  - 2i(\Psi_2 + \frac{R}{12}) 
n_{[a}\overline{m}_{b]}\, .
\ee
The right side can be simplified by expanding $\alpha$ in terms of 
$\mu$ and $V$ and using the expression (\ref{mu}) of $\mu$:
\be 
-2(\partial_{[a}  - 2\alpha_{[a})\,\beta_{b]}
= -2(\partial_{[a}  - i\nu_{[a})\,\beta_{b]} -2i (\Psi_2 + \frac{R}{12})
n_{[a}\overline{m}_{b]}
\ee
Hence we have the desired result:
\be
(\partial_{[a}  - i\nu_{[a})\,\beta_{b]} \= 0
\ee
The second equation in (\ref{beta}) follows from (\ref{F1}) and
(\ref{F3}).

Finally, transvect (\ref{F2}) with $l^a\overline{m}^b \i_C \i_D$ and
(\ref{F4}) with $l^a\overline{m}^b$. Equating the two expressions, we
conclude%
\footnote{This argument is due to Chris Beetle and Steve Fairhurst. We
are grateful to them for sharing this insight with us.}:
$\Phi_{20} \= 0$ and $\Phi_{02} \=0$.

\end{document}